\documentclass[twocolumn]{aastex63}

\usepackage{amsmath}

\shorttitle{Photometry of a new asynchronous polar and Paloma}
\shortauthors{Littlefield et al.}

\newcommand{\porb}{$P_{orb}$}
\newcommand{\pspin}{$P_{spin}$}

\newcommand{\tess}{\textit{TESS}}

\newcommand{\fullname}{SDSS~J084617.11+245344.1}
\newcommand{\igr}{IGR~J19552+0044}
\newcommand{\rxs}{1RXS~J083842.1$-$282723}
\newcommand{\ktwo}{\textit{K2}}
\newcommand{\kepler}{\textit{Kepler}}
\newcommand{\cpd}{cycles~d$^{-1}$}

\begin{document}

\title{\kepler\ \ktwo\ and \tess\ observations of two magnetic cataclysmic variables:\\The new asynchronous polar \fullname\ and Paloma  }

\author[0000-0001-7746-5795]{Colin Littlefield}
\affiliation{Department of Physics and Astronomy, University of Notre Dame, Notre Dame, IN 46556, USA}
\affiliation{Department of Astronomy, University of Washington, Seattle, WA 98195, USA}
\affiliation{Bay Area Environmental Research Institute, Moffett Field, CA 94035 USA}

\author{D. W. Hoard}
\affiliation{Department of Astronomy, University of Washington, Seattle, WA 98195, USA}
\author[0000-0003-4069-2817]{Peter Garnavich}
\affiliation{Department of Physics and Astronomy, University of Notre Dame, Notre Dame, IN 46556, USA}
\author[0000-0003-4373-7777]{Paula Szkody}
\affiliation{Department of Astronomy, University of Washington, Seattle, WA 98195, USA}
\author{Paul A. Mason}
\affiliation{New Mexico State University, MSC 3DA, Las Cruces, NM, 88003, USA}
\affiliation{Picture Rocks Observatory, 1025 S. Solano Dr. Suite D., Las Cruces, NM 88001, USA}
\author{Simone Scaringi}
\affiliation{Centre for Extragalactic Astronomy, Department of Physics, Durham University, South Road, Durham, DH1 3LE}
\author{Krystian Ilkiewicz}
\affiliation{Centre for Extragalactic Astronomy, Department of Physics, Durham University, South Road, Durham, DH1 3LE}
\affiliation{Astronomical Observatory, University of Warsaw, Al. Ujazdowskie 4, 00-478 Warszawa, Poland}
\author[0000-0001-6894-6044]{Mark R. Kennedy}
\affiliation{Department of Physics, University College Cork, Cork, Ireland}
\affiliation{Jodrell Bank Centre for Astrophysics, Department of Physics and Astronomy, The University of Manchester, M19 9PL, UK}

\author{Saul A. Rappaport}
\affiliation{Department of Physics and Kavli Institute for Astrophysics and Space Research, MIT, Cambridge, MA 02139, USA}

\author{Rahul Jayaraman}
\affiliation{Department of Physics and Kavli Institute for Astrophysics and Space Research, MIT, Cambridge, MA 02139, USA}

\correspondingauthor{Colin Littlefield}
\email{clittlef@alumni.nd.edu}

\begin{abstract}
    There have been relatively few published long-duration, uninterrupted light curves of magnetic cataclysmic variable stars in which the accreting white dwarf's rotational frequency is slightly desynchronized from the binary orbital frequency (asynchronous polars). We report \kepler\ \ktwo\ and \tess\ observations of two such systems. The first, \fullname\, was observed by the \kepler\ spacecraft for 80 days during Campaign 16 of the \ktwo\ mission, and we identify it as a new asynchronous polar with a likely 4.64-hour orbital period. This is significantly longer than any other asynchronous polar, as well as all but several synchronous polars. Its spin and orbital periods beat against each other to produce a conspicuous 6.77-day beat period, across which the system's accretion geometry gradually changes. The second system in this study, Paloma, was observed by \tess\ for one sector and was already known to be asynchronous. Until now, there had been an ambiguity in its spin period, but the \tess\ power spectrum pinpoints a spin period of 2.27~h. During the resulting 0.7~d spin-orbit beat period, the light curve phased on the spin modulation alternates between being single- and double-humped. We explore two possible explanations for this behavior: the accretion flow being diverted from one of the poles for part of the beat cycle, or an eclipse of the emitting region responsible for the second hump. 
	
\end{abstract}


\section{Introduction}

    \subsection{The three classes of magnetic cataclysmic variables}

        Cataclysmic variables (CVs) are interacting binaries in which a white dwarf (WD) accretes from a Roche-lobe-filling companion, usually an M-dwarf. If the WD possesses a significant magnetic field, the accretion flow from the donor star will be channeled onto the WD along its magnetic-field lines. The accreting matter produces a shock near the WD's surface, and the post-shock material cools by emitting a combination of X-ray bremsstrahlung and optical/near-infrared cyclotron radiation \citep{cropper}.

    \begin{deluxetable*}{ccccccc}
            \tablecaption{The confirmed asynchronous polars, J0846, and Paloma. \label{table:APs}}
            
            \tablehead{
                \colhead{Name} &
                \colhead{$P_{orb}$ (h)} &
                \colhead{$P_{spin}/P_{orb}$} &
                \colhead{$P_{beat}$ (d)} &
                \colhead{Distance (pc)} &
                \colhead{References}
                }
                
            \startdata
            \igr      & 1.39 & 0.972 & 2.04 & $165.5^{+1.9}_{-1.5}$ & \citet{tovmassian} \\
            \rxs      & 1.64 & 0.96 & 1.8 & $156.0^{+1.9}_{-2.2}$ & \citet{halpern} \\
            CD Ind    & 1.87 & 0.989 & 7.3  & $235.3^{+4.0}_{-3.2} $ & \citet{littlefield}\\
            Paloma    & 2.62 & 0.87 & 0.71 & $582^{+28}_{-20}$ & this work \\
            V1500 Cyg & 3.351 & 0.986 & 9.58 & $1570^{+270}_{-190} $ &\citet{pavlenko_v1500cyg}\\
            BY Cam    & 3.354 & 0.99  & 15  & $264.5^{+1.9}_{-1.7}$ & \citet{pavlenko_bycam}\\
            V1432 Aql & 3.366 & 1.002 & 62 & $450.\pm7$ & \citet{littlefield15}\\
            \fullname & 4.64 & 0.972 & 6.77 & $1230^{+800}_{-290}$ & this work\\
            \enddata
            
            \tablecomments{The listed distances are the geometric distances computed by \citet{BJ20} from Gaia EDR3 \citep{edr3}.}
            
        \end{deluxetable*}

        Magnetic CVs (mCVs) are typically divided into three broad categories---polars, intermediate polars, and asynchronous polars---depending on the the difference between the spin period (\pspin) of the accreting WD and the binary orbital period (\porb). In polars, the WD's magnetic field is strong enough to synchronize \pspin\ to \porb, and no accretion disk forms \citep[for a review, see][]{cropper}. Conversely, if \pspin\ is significantly shorter than \porb, the object is called an intermediate polar \citep[IP; ][]{patterson}. IPs tend to have accretion disks truncated by the WD's magnetic field, but if the WD's magnetosphere is large enough, it can prevent a disk from forming.
        
        The third category of mCVs, the asynchronous polars (APs), is comprised of systems in which \pspin\ and \porb\ differ by no more than several percent. As summarized in Table~\ref{table:APs}, only six confirmed APs have been previously reported, though the census could increase to seven, depending on how close \pspin\ and \porb\ are required to be.  APs are thought to be polars that have been temporarily desynchronized by nova eruptions. In their study of Nova Cygni 1975 (V1500~Cyg), \citet{ssl88} proposed that in the aftermath of the nova, the WD's envelope encompassed the binary, resulting in a coupling between the secondary and WD. When the WD's envelope subsequently shrank, the coupling ceased, which reduced the WD's moment of inertia and caused it to spin up, leaving V1500~Cyg in its current asynchronous state. The short-lived nature of this differential rotation is supported by the observed period-derivative trend toward synchronous rotation in all APs with a sufficiently long observational baseline to detect a change in \pspin\ \citep[e.g., as in V1500~Cyg; ][]{ss91}. The key observational distinction between IPs and APs is that in the former, asynchronous rotation is a stable equilibrium \citep{kl91, king, kw99}, while in the latter, it is not.

        Caused by the inequality of \pspin\ and \porb, the differential rotation of the WD produces a number of observable effects in both IPs and APs. The accretion flow in mCVs must become magnetically confined at some point after it leaves the secondary, and the relative orientation of the WD's magnetic field will determine both its path through the magnetosphere and the region where it accretes onto the surface of the WD. In APs and diskless IPs, the magnetosphere rotates with respect to the ballistic accretion stream; this differential rotation causes the stream to gradually plow into different regions of the WD magnetosphere. The stream's ballistic trajectory is stationary in the binary rest frame, so differential rotation occurs at the spin-orbit beat frequency of $\omega-\Omega$, where $\omega = P_{spin}^{-1}$ and $\Omega = P_{orb}^{-1}$; equivalently, $\omega-\Omega$ is the rotational frequency of the WD in the co-rotating binary rest frame. 
        
        The combination of potentially complex magnetic field structures and asynchronism has been modeled magneto-hydrodynamically \citep{zhilkin12,zhilkin16}, which suggests that pole-switching may also accompany changes between one and two accretion pole configurations. However, establishing unique magnetic field configurations from observations remains difficult.

    \begin{figure*}
        \centering
        \includegraphics[width=\textwidth]{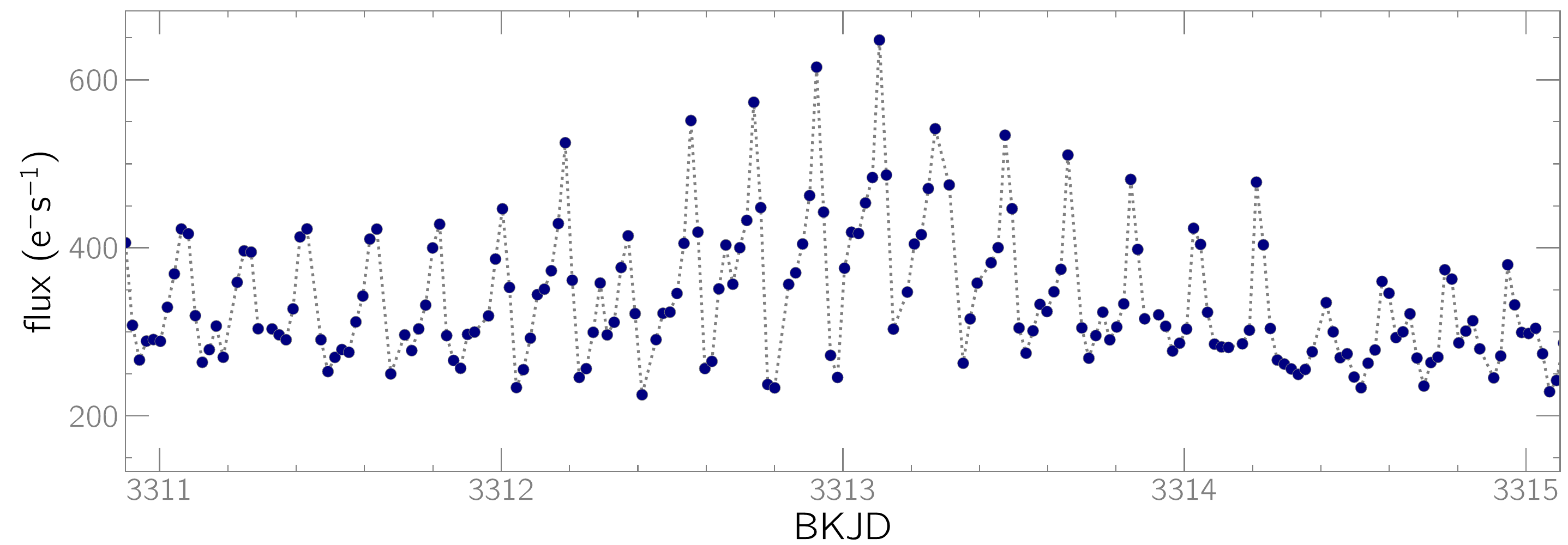}
        \includegraphics[width=\textwidth]{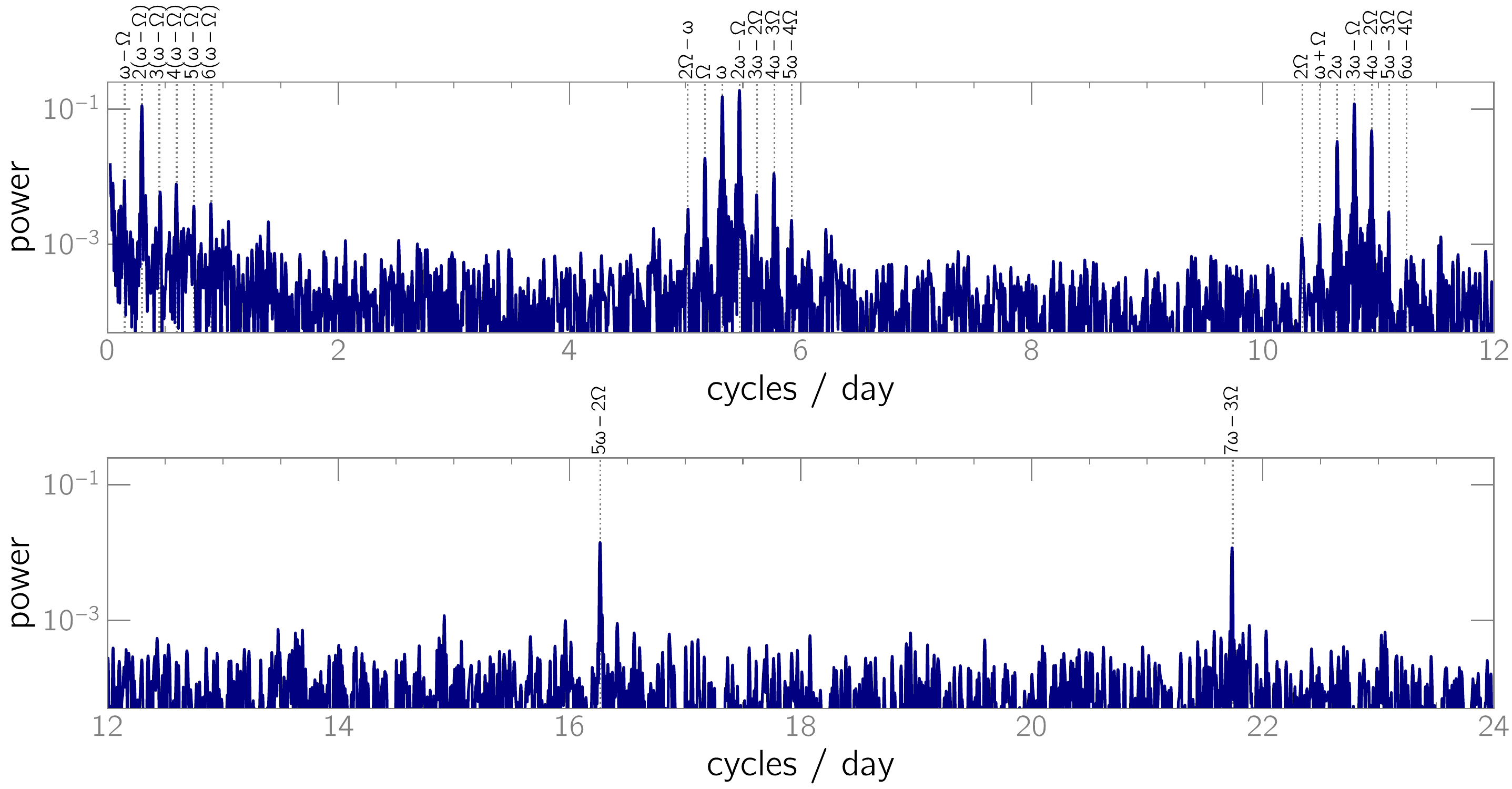}
        \caption{{\bf Top panel:} Representative $\sim$4~d segment of the \ktwo\ light curve of J0846. Each datum is a 30-minute integration, and consecutive measurements have been joined with dotted line segments to guide the eye. Flux uncertainties are negligibly small at this scale. BKJD is defined as JD$-$2454833. {\bf Bottom two panels:} Lomb-Scargle power spectrum of the \ktwo\ light curve of J0846, split into two panels to improve the visibility of major frequencies.  $\omega$ is the WD's spin frequency and $\Omega$ is the binary orbital frequency.}
        \label{fig:J0846_LC_power}
    \end{figure*}

        From an observational standpoint, the ever-changing accretion geometry of APs leads to several dramatic effects. For example, the accretion region on the WD will migrate across the WD's surface, following the footprints of whichever magnetic field lines are capturing the accretion stream at that particular time \citep{gs97}. Likewise, the bulk of the accretion flow will travel to different accretion poles during different portions of the beat cycle, and when the accretion flow switches between poles, the light curve will show a discontinuity in phase \citep{mason89}.

        \begin{figure*}
            \centering
            \includegraphics[width=\textwidth]{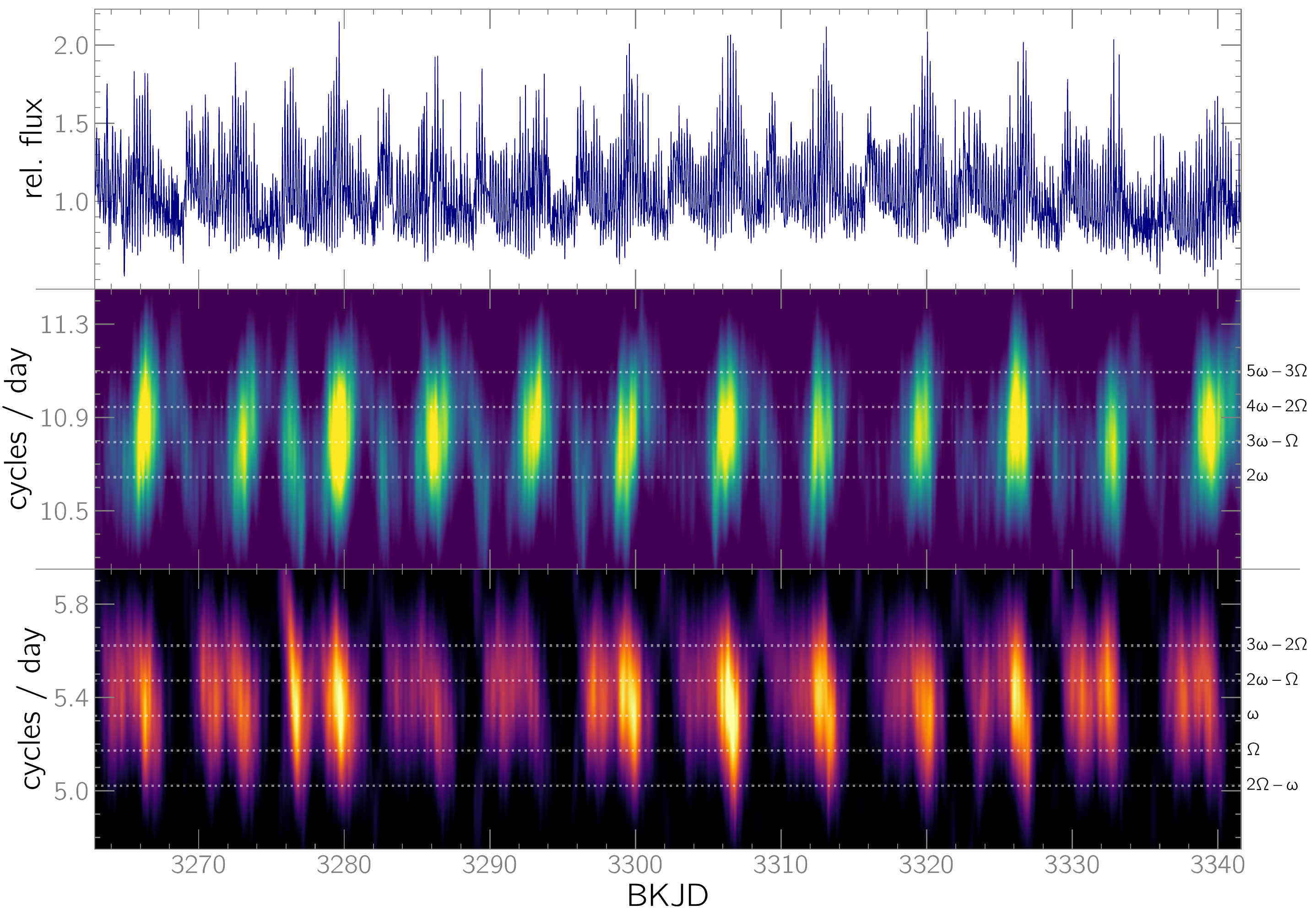}
            \caption{Light curve and 2D power spectra for J0846. BKJD is defined as BJD-2454833. The middle and lower panels use different colormaps to reflect that they have different intensity cuts. The size of the sliding window is 0.5~d. }
            \label{fig:J0846_2D_power}
        \end{figure*}
    
        \begin{figure}
            \centering
            \includegraphics[width=\columnwidth]{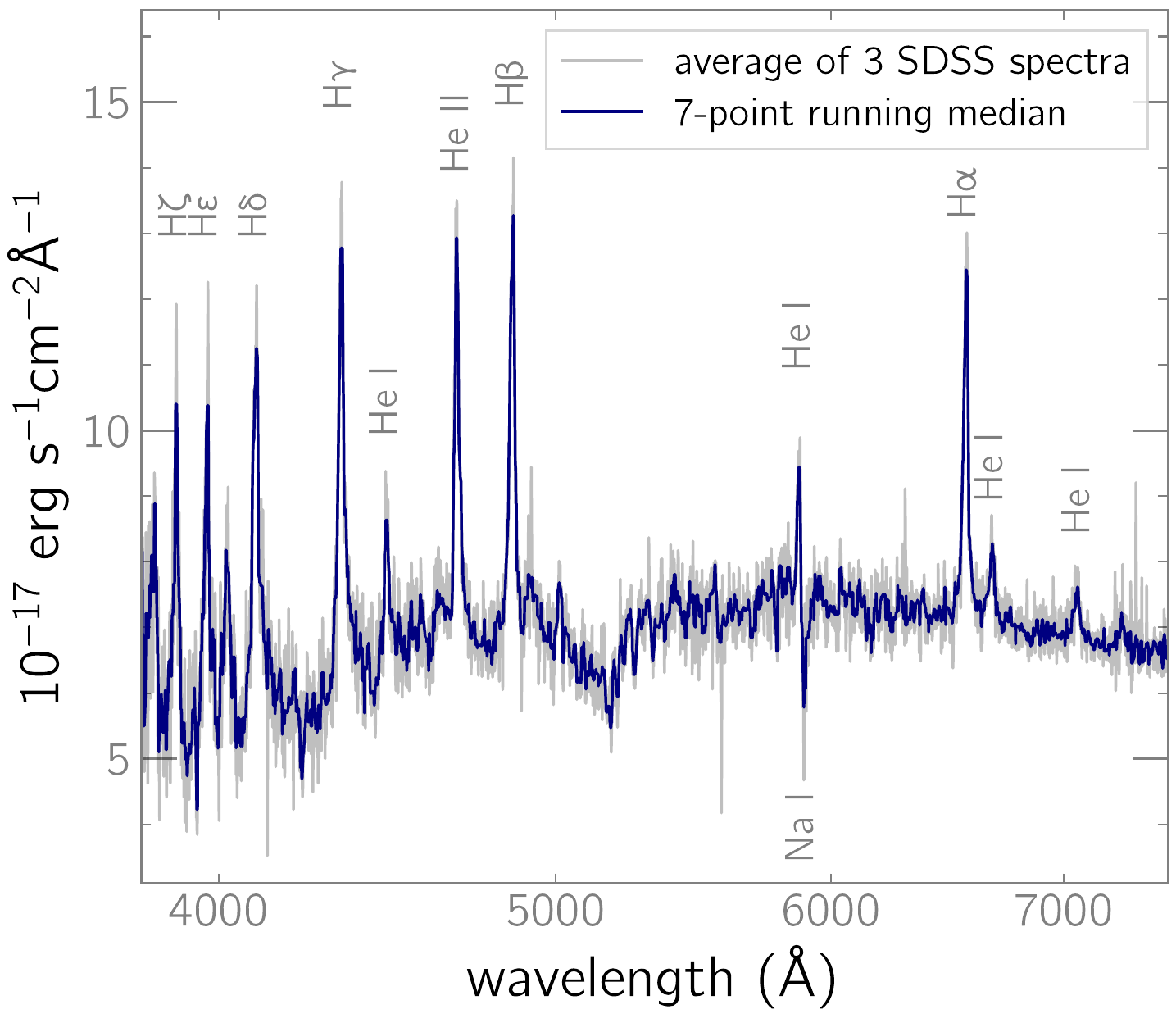}
            \caption{Average of three SDSS spectra of J0846, including one obtained during the \ktwo\ observation, showing very strong He~II~$\lambda 4686$~\AA\ emission and single-peaked emission lines. Both are commonly observed features in spectra of polars. Overall, this spectrum is similar to the one reported in \citet{szkody06}, except that He~I emission is more pronounced here. The depression in the continuum near 5200~\AA\ is suggestive of MgH absorption which, combined with the shape of the continuum and the lack of obvious VO and TiO bands, would be consistent with a late K or M0 donor star.}
            \label{fig:J0846_spectrum}
        \end{figure}

\subsection{\fullname}
      
      The cataclysmic variable \fullname\ (hereafter, J0846) has an exceedingly sparse observational history. \citet{szkody06} found that its Sloan Digital Sky Survey (SDSS) spectrum contains unusually prominent He~II $\lambda$4686\AA\ emission, a common indicator of magnetic accretion. The same study was unable to detect circular polarization in the source in a single 6000-second exposure. J0846 is listed in the Catalina Survey Periodic Variable Star Catalog with a period of 0.1827862~d, with no uncertainty specified \citep{drake}.
      
      The Gaia EDR3 \citep{gaia, edr3} distance to J0846 is $1230^{+800}_{-290}$~pc \citep{BJ20}. Its Galactic latitude of $+35.4^\circ$ therefore places it $710^{+460}_{-170}$~pc above the Galactic plane, which is significantly larger than nearly every polar \citep{beuermann21}.

\subsection{Paloma}

    The second subject of the present study, Paloma\footnote{Although CVs rarely have common names, Paloma (Spanish for ``dove'') is an exception. It acquired its name because of its chance superposition next to an unrelated, dove-shaped supernova remnant \citep{schwarz}.} (= RX J0524+42), is a rare hybrid between IPs and APs. \citet{schwarz} and \citet{joshi} published in-depth photometric and X-ray studies, respectively, but Paloma has received scant attention otherwise. \citet{schwarz} measured an orbital period of 2.62~h and constrained the spin period to be either 2.27~h or 2.43~h. These differ from the orbital period by 13\% and 7\%, respectively, so the system could be plausibly classified as either a nearly synchronous IP or a highly asynchronous AP. \citet{schwarz} also discuss the evolutionary implications of the unusual $P_{spin}/P_{orb}$ ratio, including the intriguing possibility that it is an IP evolving into a polar, a process envisioned by \citet{chanmugam}. Power spectral analysis of the X-ray light curve suggests the absence of an accretion disk \citep{joshi}. 
    
    The distance to Paloma based on Gaia EDR3 is $582^{+28}_{-20}$~pc \citep{BJ20}. Unlike J0846, it is situated very close to the Galactic plane, with a Galactic latitude of $+3.9^\circ$.

            \begin{figure*}
                \centering
                \includegraphics[width=\textwidth]{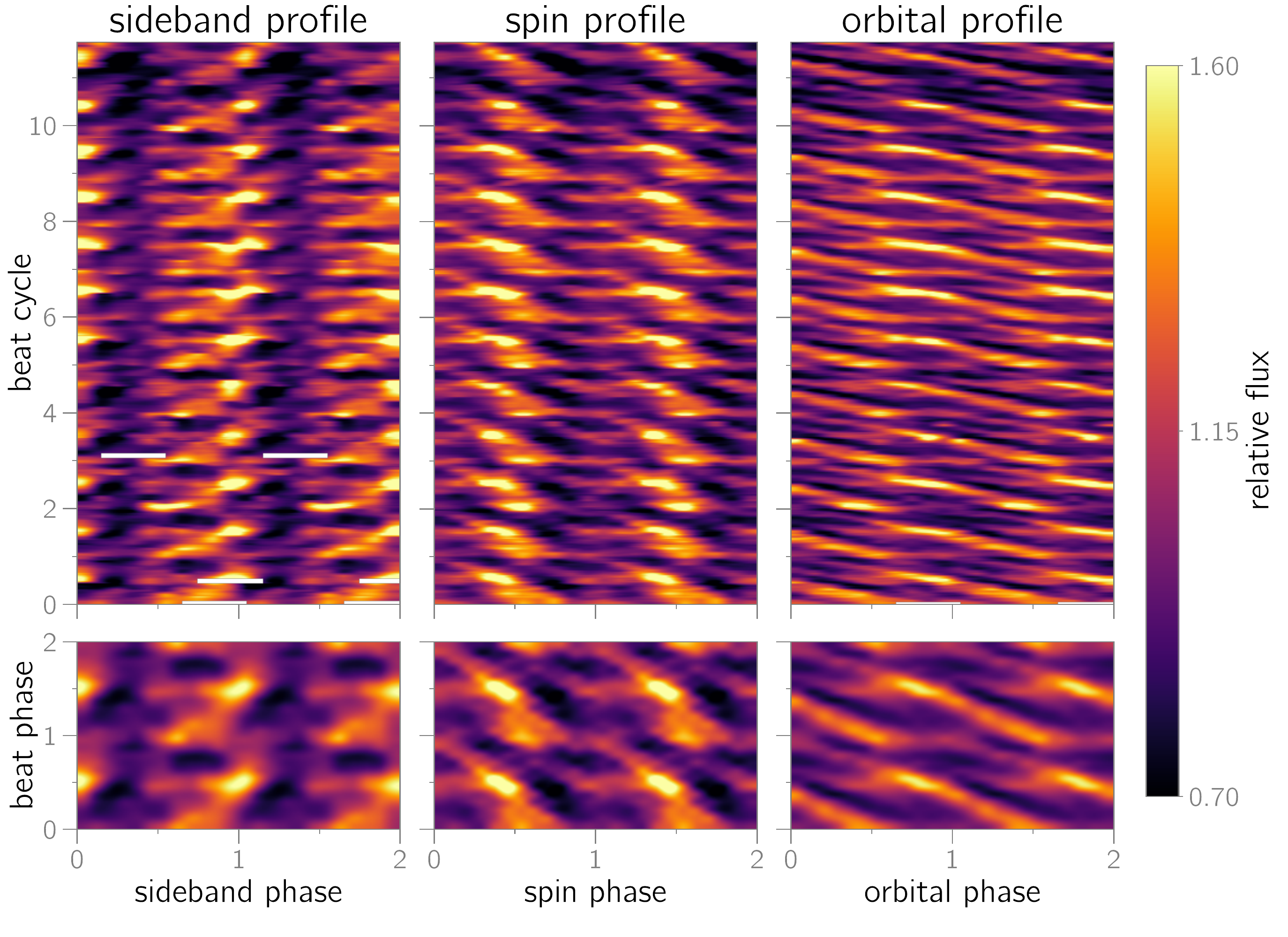}
                \caption{Two-dimensional light curves of J0846 phased to the sideband, spin, and orbital periods. Phase 0.0 is arbitrary in all panels.}
                \label{fig:2D_lightcurves}
            \end{figure*}

\section{Data}

    \subsection{The \ktwo\ observation of J0846}
    
        The \kepler\ spacecraft observed J0846 during Campaign~16 of its \ktwo\ mission between 2017~December~7 and 2018~February~25. The observations utilized the long-cadence mode, so the integration time of each datum is 30~min. 
        
        We extracted the light curve of J0846 using {\tt lightkurve}. To compensate for \kepler's well-known pointing oscillations, we chose a sufficiently large extraction aperture to encompass the full range of J0846's drift across the sensor. J0846 is situated in a sparse star field, and its signal does not suffer from serious blending.

    \subsection{The \tess\ observation of Paloma}
    
        The Transiting Exoplanet Survey Satellite (\tess) observed Paloma in its two-minute-cadence mode during Sector 19, between 2019 November 28 and 2019 December 23. The observations were uninterrupted with the exception of a day-long downlink gap in the middle of the sector. Because of the location of Paloma in a dense star field and the low angular resolution of \tess\ images, it is heavily blended with nearby sources.
        
        The \tess\ pipeline creates two versions of each two-minute-cadence light curve: simple-aperture photometry (SAP) and pre-conditioned simple-aperture photometry (PDCSAP). The PDCSAP light curve attempts to remove the effects of blending and systematic trends in the data, while the SAP light curve does not. Although this issue has not been addressed authoritatively in the context of CVs, the SAP and PDCSAP fluxes appear to show the same periodic variability, but they can differ significantly with respect to aperiodic variability. For example, the SAP light curve of TX~Col and simultaneous ground-based photometry both show an outburst, while the PDCSAP light curve does not \citep{littlefield21, rawat}.
        
        The SAP light curve of Paloma shows a gentle parabolic curvature, while the PDCSAP light curve lacks any overall trend. To determine which light curve to use, we follow the general approach of \citet{hill22} and compare both the SAP and PDCSAP light curves against simultaneous $r$-band photometry from the Zwicky Transient Facility (ZTF; \citealt{bellm2019}). We used linear regressions to compare the flux of the seven available ZTF observations with the simultaneous \tess\ observations and found that the PDCSAP light curve agreed well with the ZTF data, with a coefficient of determination of $r^2_{PDCSAP} = 0.54$. In comparison, the SAP light curve initially had $r^2_{SAP} = 0.0$, largely because of the influence of a single ZTF measurement that strongly disagreed with the trend in the SAP data. Arbitrarily removing this point resulted in $r^2_{SAP} = 0.29$, but there are no obvious indicators that that particular ZTF measurement is unreliable.
        
        Therefore, on the basis of these comparisons, we elected to use the PDCSAP light curve. We stress that unlike TX~Col, there are no astrophysically noteworthy differences between the SAP and PDSCAP light curves, so the choice between these two datasets does not significantly impact the results of our analysis. Since a few of the PDSCAP flux measurements are negative, we added an arbitrary constant offset to the PDCSAP flux.
        

    \subsection{Paloma spectra}
    
        On 2019 December 19, during the \tess\ observation of Paloma, we obtained time-resolved spectroscopy with the Large Binocular Telescope (LBT).\footnote{The LBT is an international collaboration among institutions in the United States, Italy and Germany. LBT Corporation partners are: The University of Arizona on behalf of the Arizona university system; Istituto Nazionale di Astrofisica, Italy; LBT Beteiligungsgesellschaft, Germany, representing the Max-Planck Society, the Astrophysical Institute Potsdam, and Heidelberg University; The Ohio State University, and The Research Corporation, on behalf of The University of Notre Dame, University of Minnesota and University of Virginia.} From 8:06~UT until 10:30~UT, we obtained a series of 180~s exposures with the MODS spectrographs \citep{PoggeMODS}, a 250~lines~mm$^{-1}$ grating, and a 0.8~arcsec slit aligned to the parallactic angle. During this sequence, the airmass ranged from 1.05 to 1.36. All spectra were flux-calibrated and reduced using IRAF\footnote{IRAF is distributed by the National Optical Astronomy Observatory, which is operated by the Association of Universities for Research in Astronomy (AURA) under a cooperative agreement with the National Science Foundation.} standard procedures.
        
        The LBT spectra are extremely complex and will be the subject of a dedicated spectroscopic paper. As a result, in this study, we rely upon them sparingly (primarily to establish an orbital ephemeris in order to phase the photometry to the binary orbit).

\section{Analysis of the new asynchronous polar J0846}

    \subsection{Light curve}
    
        The top panels of Figs.~\ref{fig:J0846_LC_power}~and~\ref{fig:J0846_2D_power} show a representative segment of the \ktwo\ light curve of J0846 and its full light curve, respectively. Despite the 30-min cadence of the observations, it is obvious that J0846 has large-amplitude variability on timescales shorter than the observational cadence. At times, the flux doubles in the span of several hours and is halved in even less time, giving the light curve a jagged appearance. The profiles of individual photometric maxima, with their large amplitudes and rapid changes, are typical for a polar. However, unlike normal polars, both the amplitude and shape of the maxima gradually evolve over a 6.7-d period before returning to their original appearance. This highly periodic and well-defined modulation of the short-term variability is the distinguishing property of J0846's light curve and provides compelling evidence that it is an AP. As we explain in detail in Sec.~\ref{sec:frequency_ID}, we identify the 6.7-d period in J0846 as the beat between its likely spin ($\omega$=5.32~\cpd) and orbital ($\Omega$=5.17~\cpd) frequencies.
        
        The changes in the light curve across the beat period result in a rich power spectrum (Fig.~\ref{fig:J0846_LC_power}, lower panels) containing $\omega$, $\Omega$, and numerous sidebands and harmonics thereof. The time-resolved power spectrum (Fig.~\ref{fig:J0846_2D_power}) shows that the power spectrum varies cyclically at the beat period. Here again, this behavior is expected in an AP and has been observed in \tess\ observations of the AP CD~Ind \citep{hakala, littlefield, mason20}.

        Our classification of J0846 as an AP is supported by the previously reported observations of the system. As we noted earlier, an SDSS spectrum obtained in 2004 and published in \citet{szkody06} was consistent with J0846 being a polar. Fig.~\ref{fig:J0846_spectrum}, which averages the previously reported SDSS spectrum with two additional spectra obtained in 2018 and 2019, confirms that He~II~$\lambda$4686\AA\ is of comparable strength to H$\beta$ and that the emission lines are single-peaked, properties that are commonly observed in polars.\footnote{To improve legibility, the individual spectra are not shown in Fig.~\ref{fig:J0846_spectrum}, but except for changes in the strength of the He~I emission, the three spectra were largely similar.} 
        Although \citet{szkody06} did not detect circular polarization in a single 6000~s interval, this timespan covered significantly less than half of one cycle of the photometric variations in Figures~\ref{fig:J0846_LC_power}~and~\ref{fig:J0846_2D_power}. Inopportune sampling could therefore explain the absence of circular polarization in that observation.

    \subsection{Interpreting J0846's power spectrum}
    
    \subsubsection{Theoretical considerations for AP power spectra}

        \begin{deluxetable}{ccc}
                \tablecaption{J0846 frequency identifications.\label{table:frequencies}}
                \tablehead{\colhead{Frequency} & \colhead{Case 1} &             \colhead{Case 2} }
                \startdata
                $\Omega$         & 5.17 \cpd & 5.32 \cpd \\
                $\omega$         & 5.32 \cpd & 5.47 \cpd \\
                $2\omega-\Omega$ & 5.47 \cpd & 5.62 \cpd \\
                \enddata
                \tablecomments{We argue in favor of Case 1, but Case 2 is possible.}

        \end{deluxetable}

            \begin{figure*}
                \centering
                \includegraphics[width=\textwidth]{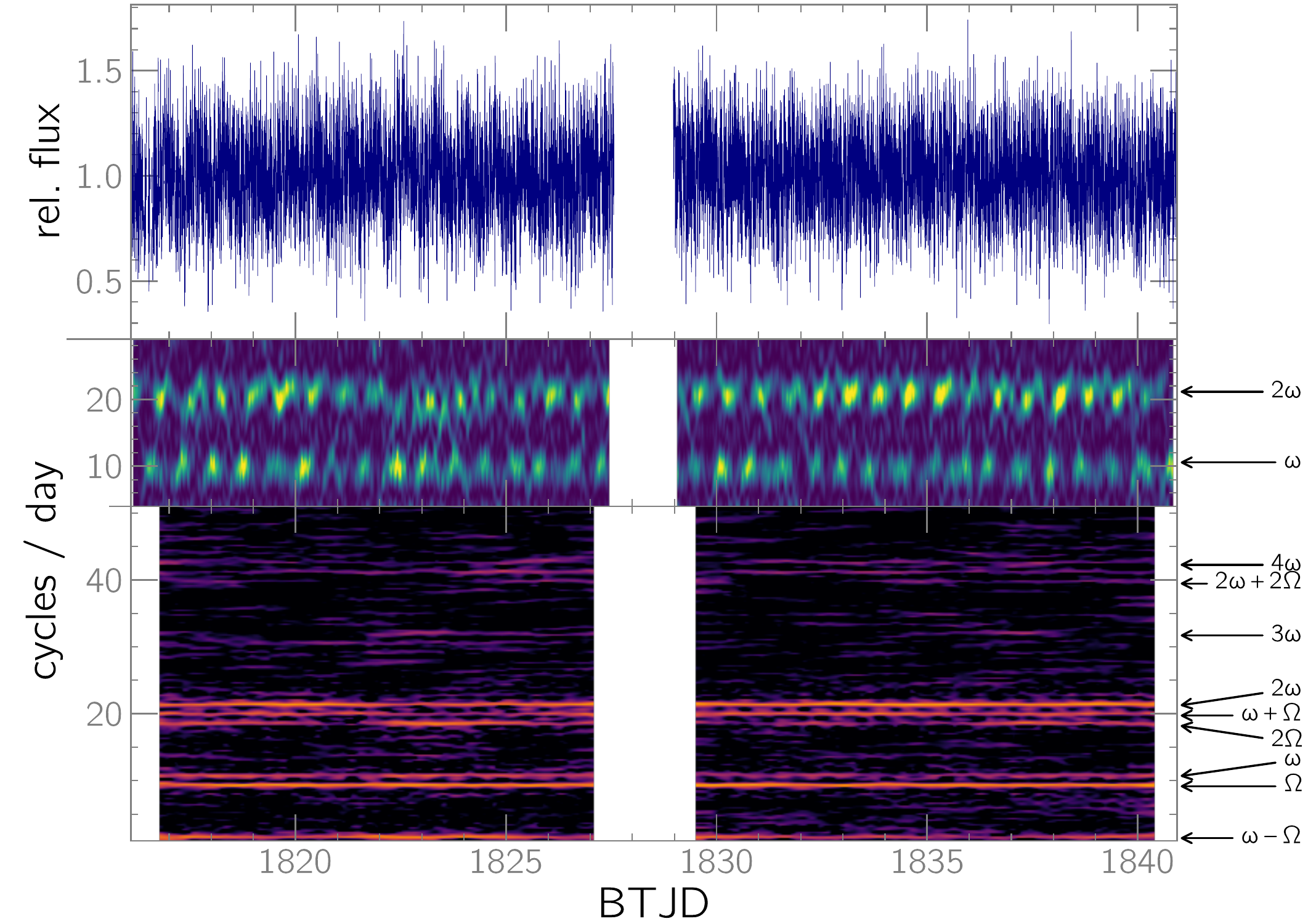}
                \caption{{\bf Top:} The \tess\ light curve of Paloma. BTJD is defined as BJD - 2457000. {\bf Middle:} Two-dimensional power spectrum with a 0.25-d sliding window, with linear intensity scaling. Across the 0.7-d beat cycle, the light curve alternates between a single- and double-humped profile, causing power to shift cyclically between $\omega$ and $2\omega$. {\bf Bottom:} Two-dimensional power spectrum with a 2-d sliding window and logarithmic scaling. The larger window size offers improved frequency resolution at the expense of completely concealing the changes evident in the middle panel.}
                \label{fig:Paloma_trailed_power}
            \end{figure*}

            \begin{figure*}
                \centering
                \includegraphics[width=\textwidth]{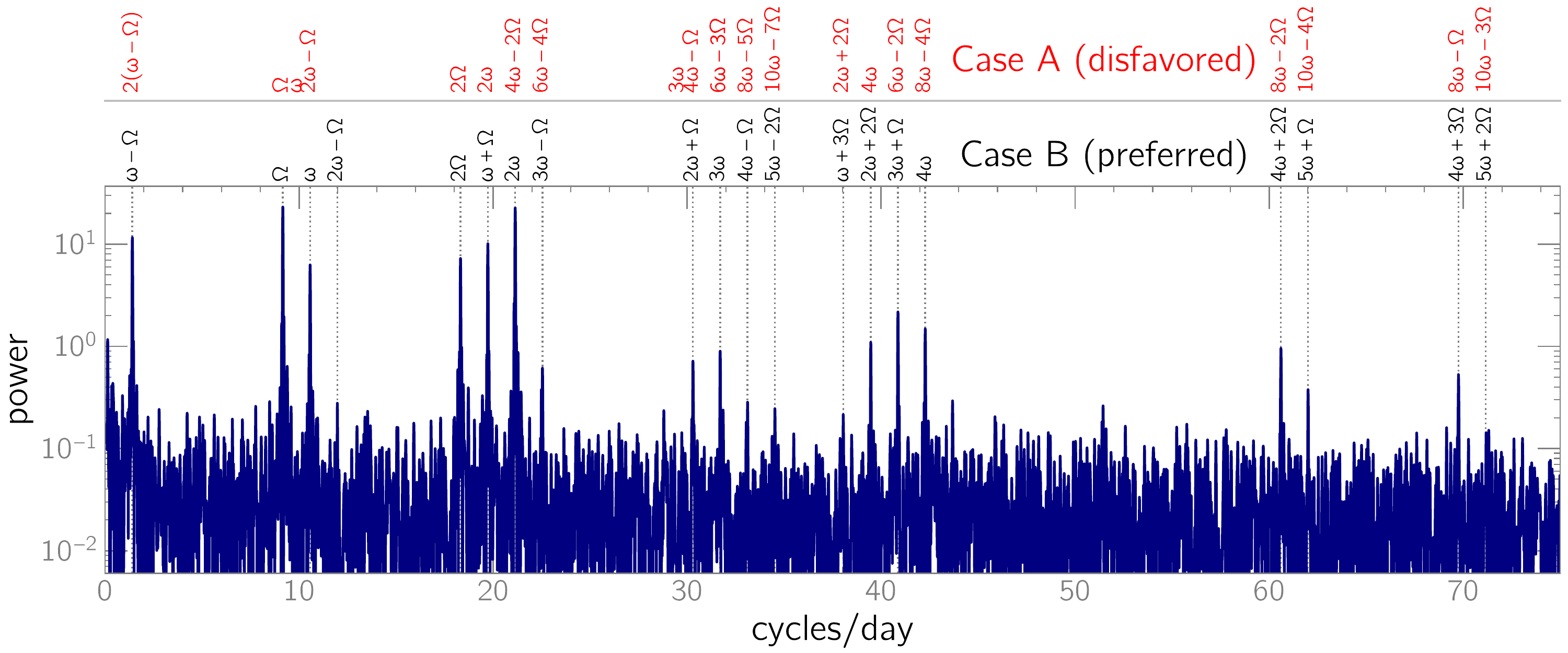}
                \caption{Lomb-Scargle power spectrum of the \tess\ light curve of Paloma, with the two sets of frequency identifications proposed by \citet{schwarz}. The usage of ``Case~A'' and ``Case~B'' matches that of \citet{schwarz}. For the Case~A identifications, there would be no appreciable signal at either $\omega$ or $\omega-\Omega$. For reasons discussed in the text, we prefer the ``Case B'' identifications.
                \label{fig:Paloma_power} }
            \end{figure*}

        Even with a long, uninterrupted light curve, the identification of the spin and orbital frequencies in APs is fraught with difficulties not otherwise encountered in the study of CVs.
    

            With synchronous polars, the accretion region is expected to be stationary at a fixed mass-transfer rate, so for purposes of measuring the WD's spin period, it is often treated as a fiducial marker of the star's rotation. In APs, however, the accretion region moves across the surface of the WD and even jumps between magnetic poles, making the photometric modulation of the accretion region an unreliable indicator of the spin period. 
            
            Indeed, by causing large phase shifts in the light curve, the movement of the accretion region wreaks havoc on the power spectrum. \citet{wk92} predicted that in X-ray light curves of IPs, pole switching could cause the dominant frequency in the power spectrum to be $2\omega-\Omega$, even if the light curve is modulated at $\omega$ between the pole switches \citep{mason20}. The widely-used Lomb-Scargle periodogram (\citealt{lomb1976}; \citealt{scargle1982}), along with other common period-finding algorithms, presume that a signal does not experience these large, regular phase jumps, and if pole switching is present in a light curve, these algorithms will be biased towards the identification of a period that forces a signal to remain as in-phase as possible
            \citep[Sec.~4.1 in][]{littlefield}.

            \citet{mason95} and \citet{mason98} extended the rationale of \citet{wk92} to optical observations of APs and identified the $2\omega - \Omega$ sideband as the strongest signal in the power spectrum of BY~Cam. In a similar vein, \citet{littlefield} concluded from the \tess\ light curve of CD~Ind that the long-accepted identification of the spin frequency in that system is actually $2\omega-\Omega$ (although this proposal awaits independent spectroscopic confirmation). 
            
            There is yet another complication: even between pole switches, the accretion region is expected to move longitudinally across the surface of the WD \citep{gs97}. The asynchronous rotation of the WD, with respect to the binary, causes the accretion stream to thread onto a continuously changing ensemble of magnetic field lines, each of which channels material onto different points along the WD's surface. Consequently, in an AP, the interval between the accretion region's crossings of the WD's meridian can differ by several percent from the true WD rotational period \citep{gs97}. 
            
            Although accurately identifying $\omega$ and $\Omega$ from photometry alone is therefore a challenging affair, it is comparatively easy to identify their beat frequency ($\omega - \Omega$). Even if $\omega-\Omega$ is not directly visible in the power spectrum, it will be observable as the spacing between sideband frequencies of $\omega$ and $\Omega$.

    \begin{figure*}
        \centering
        \includegraphics[width=\textwidth]{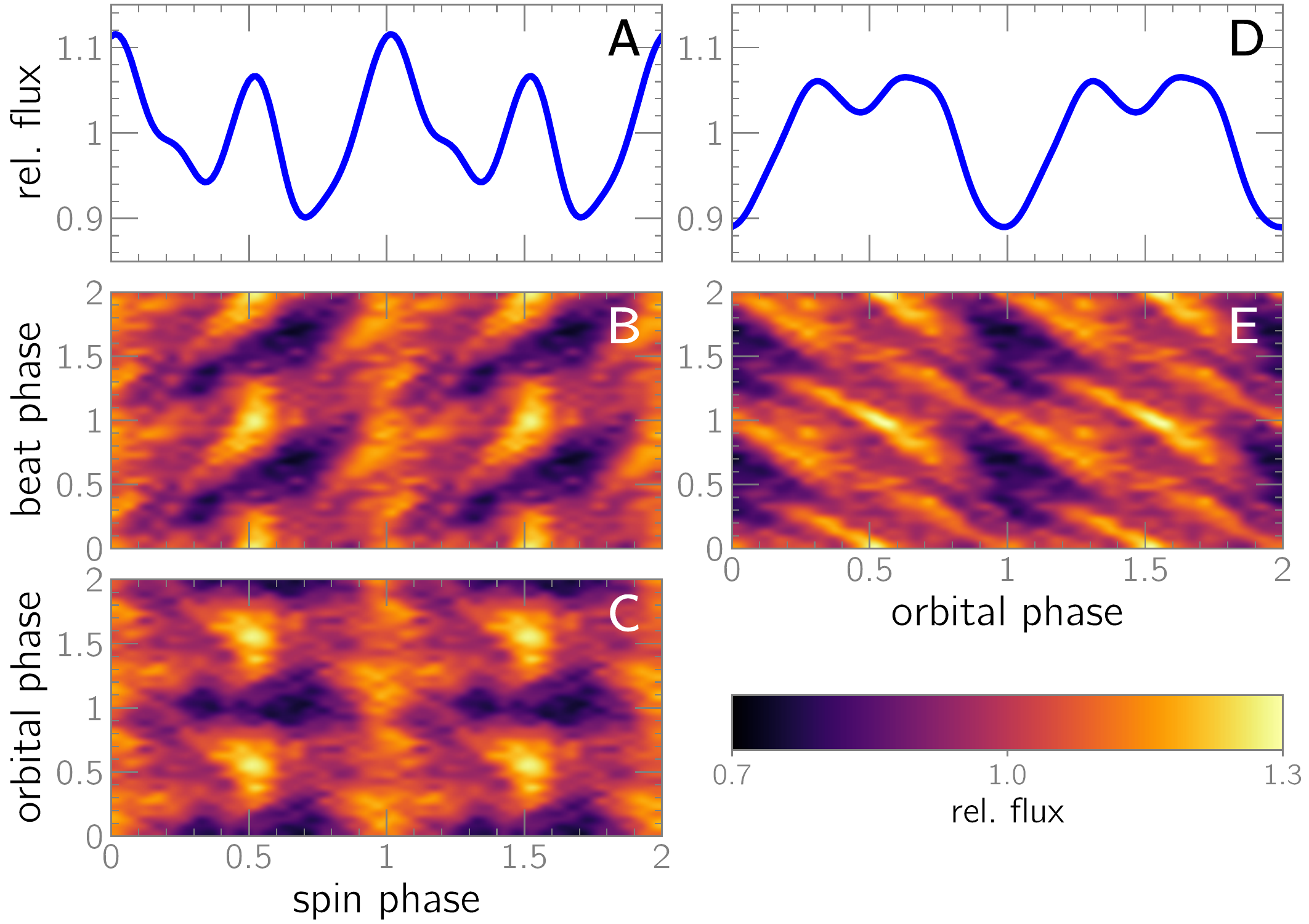}
        \caption{Interdependence of Paloma's spin and orbital profiles across the beat cycle (relative to $T_0,beat [BJD] = 2458836.1495$). Panels A-C share the same $x$-axis (spin phase relative to $T_0,spin [BJD] =  2458836.9065$), while panels D and E show the orbital phase on a common $x$ axis. Panels A and D present six-harmonic Fourier-series representations of the average spin and orbital profiles across the entire \tess\ observation. Horizontal slices through panels B and C yield the spin profile that would be observed if the beat and orbital phases, respectively, could be held constant, and Panel C shows that the secondary spin maximum disappears at inferior conjunction. Similarly, panel E shows the evolution of the orbital profile across the beat cycle. Panel C establishes that the spin profile changes significantly near inferior conjunction. 
        }
        \label{fig:paloma_2D_lightcurves}
    \end{figure*}

    \subsubsection{Frequency identifications in J0846}
    \label{sec:frequency_ID}
        
            With these considerations in mind, we turn to the power spectrum of J0846 and examine two sets of possible frequency identifications in the observed power spectrum of J0846.
            
            The unusually rich power spectrum of J0846 (bottom panels of Fig.~\ref{fig:J0846_LC_power}) bears many similarities with the \tess\ power spectrum of CD~Ind and is consistent with J0846 being an AP. The major signals in the power spectrum are clustered in three groups, and the signal with the most power occurs at a frequency of 5.47~\cpd. This is the same frequency measured by the \citet{drake} pipeline from survey photometry. Nearby at 5.32~\cpd\ is another major signal. At lower frequencies, there is a family of six harmonically related signals, with the fundamental being 0.15~\cpd; this is also the spacing between the frequencies in the other two clusters of signals. 
            
            The power spectrum is amenable to two sets of frequency identifications, and while both agree that the beat frequency ($\omega-\Omega$) is 0.15~\cpd, they diverge on the correct identifications of $\omega$ and $\Omega$. Following \citet{mason95} and \citet{mason98}, we propose that the highest-amplitude signal (5.47~\cpd) is the $2\omega-\Omega$ sideband. In this scenario, which we shall refer to as Case~1, \pspin=5.32~\cpd\ and \porb=5.17~\cpd. The phased light curves based on the Case~1 identifications are presented in Fig.~\ref{fig:2D_lightcurves}. The $2\omega-\Omega$ sideband profile shows the behavior qualitatively explained by \citet{littlefield} for CD~Ind; when phased to this frequency, the pulses remain comparatively in phase throughout the observation. \citet{littlefield} contended that this is due to a bias of frequency-analysis algorithms. The spin-phased profiles, conversely, show evidence of discrete, variable accretion regions on opposite sides of the WD. 
  
            There is an additional set of plausible frequency identifications in which the dominant signal in the power spectrum would be the spin frequency, such that $\omega=5.47$~\cpd. In this scenario, which we call Case~2, $\Omega=5.32$~\cpd. Returning to the phased light curves in Fig.~\ref{fig:2D_lightcurves}, the nominal sideband-phased and spin-phased light curves from Case~1 would actually be the spin-phased and orbit-phased light curves, respectively, in Case~2.
            
            We summarize both sets of frequency identifications in Table~\ref{table:frequencies}. While we favor Case~1, the proper identification of the orbital period can be conclusively ascertained with time-series spectroscopy of the secondary. An undisputed orbital period, in combination with the 6.7~d beat period, would also eliminate any remaining ambiguity surrounding the spin period.
            
            In either set of frequency identifications, J0846 would have an unusually long orbital period for a polar. At the time of writing, the International Variable Star Index (VSX) catalog contains 145 confirmed or candidate polars;\footnote{We exclude the object CG~X-1 from this group. CG~X-1 was formerly considered a candidate polar and remains identified as such in the VSX at the time of writing. However, \citet{esposito} and \citet{qiu} reclassified it as an extragalactic high-mass X-ray binary.} if the orbital period of J0846 is 4.64~h, as we have argued, only three systems (V895~Cen, V1309~Ori, and V479~And) would have longer orbital periods. That census would increase to only four polars (with AI~Tri being the fourth) if J0846's orbital period is instead 4.51~h, as it would be in the second, disfavored set of frequency identifications.

        \subsection{Accretion geometry} \label{sec:geometry}
            
            Assuming that Case~1 correctly identifies the spin period, the profile of the spin pulse across the beat cycle is extremely intricate. The spin profile often resembles that of a synchronous polar across short intervals of the beat cycle, particularly during beat phases 0.0-0.25 (where $T_{0,beat}$ is arbitrarily defined as BJD=2458095.4882). During the next quarter of the beat cycle, the profile develops a plateau, while its peak becomes sharp and narrow, with a conspicuous dip after the pulse maximum. For much of the remainder of the beat cycle, the pulse profile becomes comparatively ill-defined, particularly at beat phase 0.8. Nevertheless, the main accretion region appears to be active for well over half of the beat cycle.
            
            These behaviors are difficult to reconcile with a centered, dipolar field. Such a configuration would be expected to result in diametrically opposed accretion regions on opposite sides of the WD, with each pole accreting during opposite halves of the beat cycle. There is very clearly a dominant accretion region near spin phase 0.5, but there is also a signal near spin phase 1.0 at two different points in the beat cycle. Due to the 30-min cadence of the observations, we cannot confidently discern whether this is a second accretion spot or simply an evolution of the photometric profile of the main accretion spot. The latter might occur as a result of the migration of the accretion region, both in longitude and latitude, across the beat cycle, as first described by \citet{gs97} in a different AP, V1432~Aql.
            
            Pole-switching in J0846 is much less pronounced than it is in the first \tess\ observation of CD~Ind \citep{hakala, littlefield, mason20}. In the CD~Ind light curve, there was a conspicuous jump in phase, as well as a change in the pulse profile, whenever the accretion flow switched between magnetic poles  \citep{littlefield}. In J0846, the pulse profile of the main accretion region experiences obvious changes, but it never switches off in the same manner as CD~Ind. However, the poor phase resolution of the \ktwo\ light curve means that any single rotational cycle is sampled fewer than 10 times, whereas CD~Ind's spin profile was much more favorably sampled by \tess.

    \subsection{Comparing J0846 to other APs}
    
        As summarized in Table~\ref{table:APs}, J0846 is either the seventh or the eighth AP (depending on whether Paloma is also classified as such) and it joins six other APs for which the condition $|(P_{spin} - P_{orb}) / P_{orb} | \lesssim4\%$ holds true. Only Paloma, for which $|(P_{spin} - P_{orb}) / P_{orb} |  = 0.13$, fails to satisfy this requirement; its comparatively high level of desychronization makes its inclusion in Table~\ref{table:APs} debatable. Regardless of Paloma's classification, J0846 has by far the longest orbital period of the systems listed in Table~\ref{table:APs}.
        
        Of these systems, V1432~Aql is a clear outlier, based on its very small level of asynchronism, its eclipsing nature, and the fact that it alone has $P_{spin} > P_{orb}$. On that final point, however, \citet{wang2020} used a new power-spectral modeling technique to propose new identifications of the orbital frequencies of CD~Ind and BY~Cam, and they argued that $P_{spin} > P_{orb}$ in these two systems, too.\footnote{The final paragraph in Sec.~4.2 of \citet{littlefield} discusses circumstantial evidence against this particular re-identification for CD~Ind.} The \citet{wang2020} proposal can be tested conclusively by measuring \porb\ from the radial-velocity variations of the donor star in each system; unlike the complex photometric variations, the orbital motion of the secondary must, by definition, occur at \porb.

\section{Paloma}

        \begin{figure*}
        \centering
        \includegraphics[width=\textwidth]{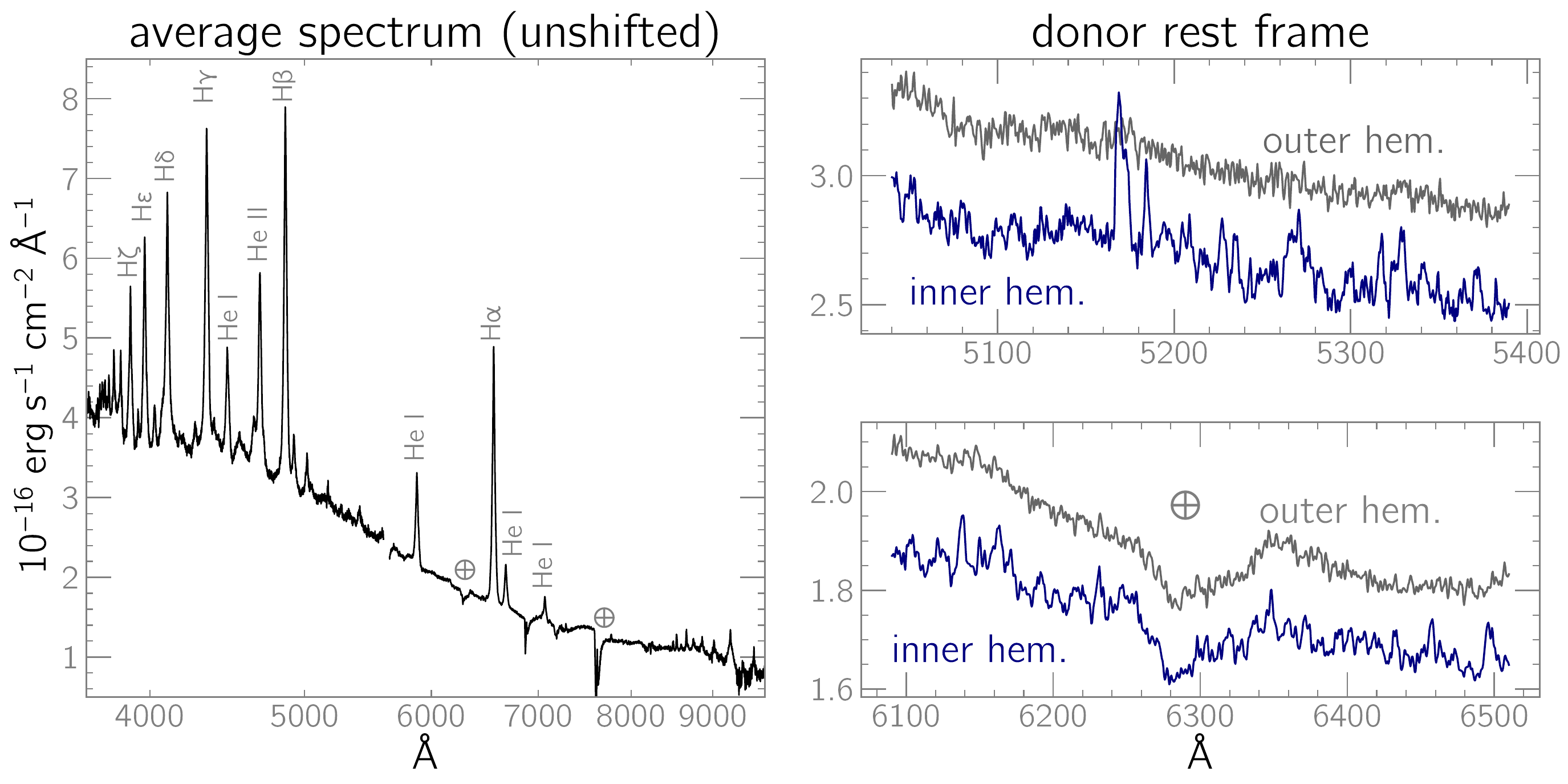}
        \caption{{\bf Left:} Average spectrum of Paloma, without any velocity shift applied. Telluric features have not been removed. {\bf Right panels:} Segments of the average spectra in the donor star's rest frame, showing a forest of weak metal emission lines from the inner hemisphere. The velocity-corrected spectra are based on the measured semi-amplitudes of 220~km~s$^{-1}$ and 310~km~s$^{-1}$ for the secondary's inner and outer hemispheres, respectively.
        }
        \label{fig:spectrum}
    \end{figure*}

    \subsection{Frequency identifications}
    
        Paloma is a faint and blended source in the TESS data, but its complex variability is visible in Fig.~\ref{fig:Paloma_trailed_power}. The sliding-window size of 6~h in the two-dimensional power spectrum in the middle panel of Fig.~\ref{fig:Paloma_trailed_power} captures the cyclical transfer of power between $\omega$ and $2\omega$ across the 0.7-d beat cycle. 
        
        The power spectrum for the full dataset (Fig.~\ref{fig:Paloma_power}) provides an opportunity to resolve the long-standing ambiguity concerning the correct identification of the spin period. \citet{schwarz} proposed and carefully justified two possible sets of frequency identifications, and with the \tess\ light curve, we can determine which is correct. Since the spectroscopic orbital frequency is unambiguously $\Omega=9.1$~\cpd\ \citep{schwarz}, there are only two plausible identifications of Paloma's signal at 10.5~\cpd: the $2\omega - \Omega$ sideband or the spin frequency $\omega$. \citet{schwarz} refer to these two scenarios as Case A and Case B, respectively. The X-ray study presented by \citet{joshi} argued for Case B, and we concur.

        In the \citet{schwarz} Case A, the identification of the $2\omega - \Omega$ sideband would require the true spin frequency $\omega$ to be equidistant between $\Omega=9.1$~\cpd\ and $2\omega - \Omega=10.5$~\cpd. Although there is a signal near this frequency in the \citet{schwarz} power spectra, there is none in either the \tess\ or \citet{joshi} power spectra. Case A further demands a beat frequency of 0.7~\cpd. The \tess\ power spectrum shows no significant power at this frequency. Moreover, for Case A to be correct, all power at the fundamental spin and beat frequencies would need to be shifted into harmonics, a scenario that is unlikely.

    Conversely, in the \citet{schwarz} Case B identifications, the frequency at 10.5~\cpd\ is $\omega$, resulting in a beat frequency $\omega - \Omega$ of 1.4~\cpd. The \tess\ power spectrum contains a very strong signal at precisely that frequency. There would also be significant signals at $\omega$ and its next three harmonics---unlike Case A, where only the $2\omega$ and $4\omega$ harmonics would have significant power. While the Case B identifications would require there to be negligible power at $2\omega - \Omega$, the appearance of this frequency in power spectra depends on the orbital inclination and the colatitude of the accretion region; it is not expected to be universally present in diskless accretors \citep{wang2020}.  
  
    We therefore agree with \citet{joshi} that Case B from \citet{schwarz}, in which $\omega-\Omega = 1.4$~\cpd\ and $\omega =10.5$~\cpd, is the correct set of frequency identifications. 
    
    A remaining loose thread from this discussion is the nature of the signal detected by \citet{schwarz} at 9.87~\cpd, the putative spin frequency in their Case A. \citet{schwarz} pointed out that in Case B, this would be the first subharmonic of the $\omega+\Omega$ sideband---\textit{i.e.}, 9.87~\cpd\ = ($\omega+\Omega)/2$. Subharmonics are not expected to be present in a Lomb-Scargle power spectrum like ours, but \citet{schwarz} used the analysis-of-variance (AoV) algorithm to compute their power spectra. One property of AoV power spectra is that they can contain subharmonics of signals, and we agree with \citet{schwarz} that this is a likely explanation for the signal at 9.87~\cpd\ in their AoV power spectrum.

    \subsection{The orbital-phase dependence of the spin pulse}

    The unbinned light curve of Paloma is rather noisy, but because the spin and orbital frequencies are known, we can phase-average the light curve to improve the signal-to-noise ratio. Fig.~\ref{fig:paloma_2D_lightcurves} reveals the complex interplay between the spin and orbital profiles throughout the \tess\ observation. The spin profile shows two distinct maxima, separated in phase by 0.5 rotational cycles. Interestingly, the secondary maximum, which occurs at spin phase 0.5, is present for only part of the beat cycle and is not visible when it coincides with the secondary's inferior conjunction (the epoch of which is measured in Sec.~\ref{sec:ephem}). In contrast, the primary spin maximum (spin phase 0.0) is present throughout the light curve and shows very little dependence on the orbital or spin phases. Because the amplitude of the secondary spin pulse is so strongly modulated across the beat cycle, we define a reference epoch ($T_0,beat [BJD] = 2458836.1495$) such that the secondary maximum attains its maximum amplitude at beat phase 0.0, and we use this definition when phase-folding data in Fig.~\ref{fig:paloma_2D_lightcurves}.

    While the orbital profile (Fig.~\ref{fig:paloma_2D_lightcurves}, panel~E) is not nearly as intricate as the spin profile, it shows a wide dip at inferior conjunction. The structure of this profile suggests that the secondary's inner hemisphere contributes significantly to the \tess\ light curve and that the dip occurs when the inner hemisphere is mostly blocked by the secondary's cool backside.
    
    There are two scenarios that could account for the behavior in Fig.~\ref{fig:paloma_2D_lightcurves}: pole switching and a grazing eclipse of one of the emitting regions. We shall consider the strengths and weaknesses of each hypothesis separately.
    
    \subsubsection{Scenario 1: pole switching}
    
    In the pole switching scenario, accretion onto one of the poles ceases for half of the beat cycle, while the other pole accretes continuously. The preference for accretion onto one of the poles would require the magnetic-field topology to be more complex than a simple, centered dipole. One strength of this explanation is that it is consistent with the power spectral evidence (both here and in \citealt{joshi}) that Paloma is a diskless IP. In the absence of a disk, one or more magnetic poles can be temporarily and periodically starved of a matter supply. Conversely, in a disk-fed IP, the inner rim of the disk provides a reservoir of material for both accretion regions, independent of the WD's rotation.
    
    \subsubsection{Scenario 2: a grazing eclipse}
    
    An alternative explanation is that the spin pulse is intrinsically the same across the observations but is extrinsically altered by an eclipse. Figure~\ref{fig:paloma_2D_lightcurves} establishes that the pulse from the second pole disappears when it coincides with the secondary's inferior conjunction. This phasing is exactly what is expected of a grazing eclipse by the donor star. 
    
    The chief difficulty with this scenario, however, is the absence of eclipses in the \citet{joshi} X-ray observations, which covered a full beat cycle. X-rays in IPs are emitted from a post-shock region just above the WD's photosphere, so an eclipse of the WD will produce sharp, energy-independent dips every orbital cycle. Thus, the non-detection of such a feature is very strong evidence that the WD itself is not eclipsed by the secondary. However, the eclipse interpretation of Fig.~\ref{fig:paloma_2D_lightcurves} nevertheless remains tenable. This is because the optical spin pulse might be produced in extended accretion curtains away from the WD, as has been observed for, e.g., FO~Aqr \citep{beardmore98}. If one of these curtains is blocked by the secondary at inferior conjunction, it could easily explain the disappearance of one of the maxima of the spin profile in Fig.~\ref{fig:paloma_2D_lightcurves} without a corresponding eclipse at X-ray energies.

    The available evidence does not offer an obvious answer as to which scenario (if either) is correct. However, it is difficult to dismiss as a coincidence the fact that one of the spin maxima disappears only when it is observed at inferior conjunction, and it is this factor that leads us to tentatively favor the eclipse interpretation. Spectroscopic observations of the eclipse-like feature might offer a more definitive answer, as the eclipse of an accretion curtain should produce a concomitant weakening of the high-velocity components of the H and He emission lines.

    \begin{figure}
        \centering
        \includegraphics[width=\columnwidth]{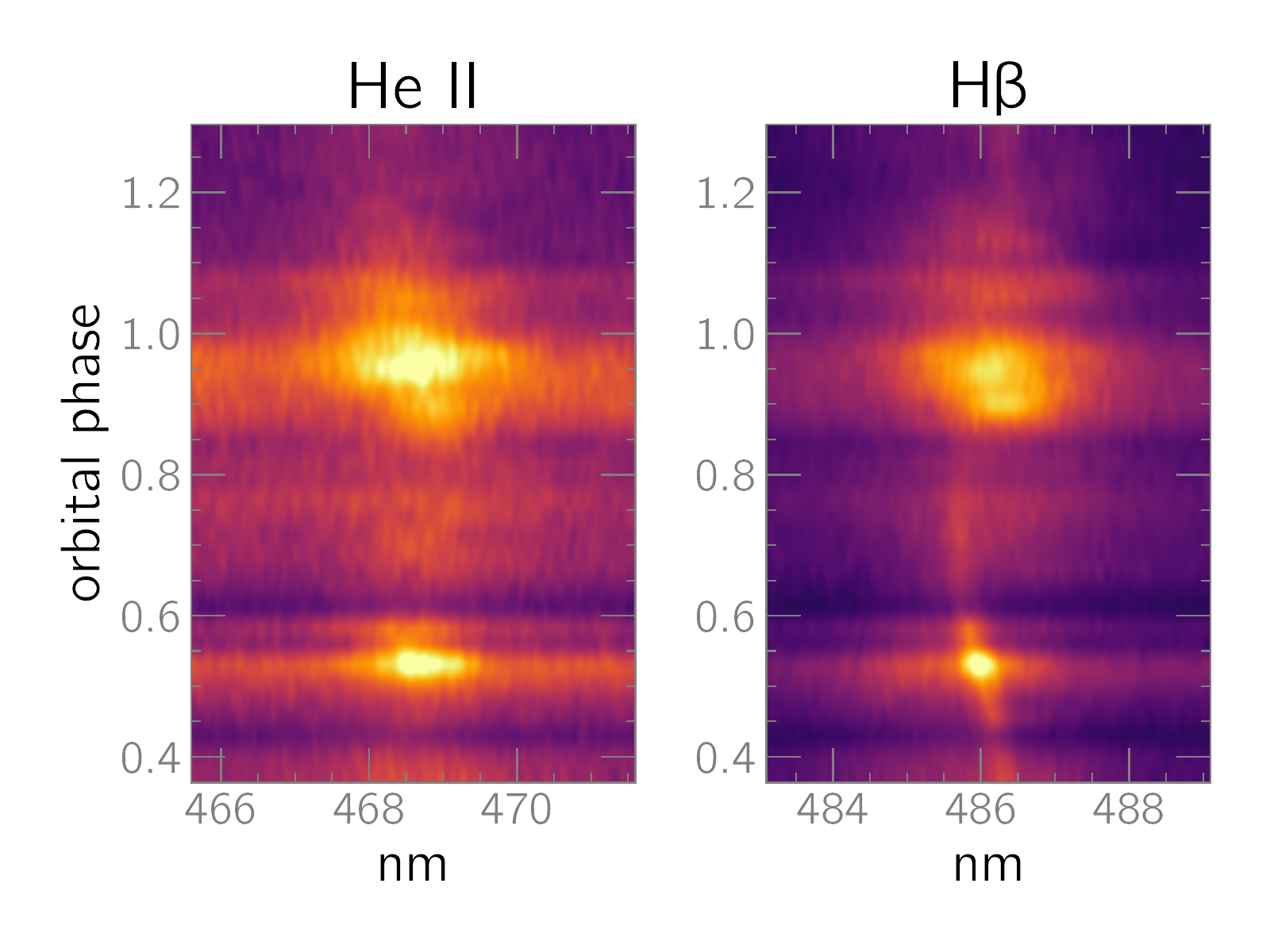}
        \includegraphics[width=\columnwidth]{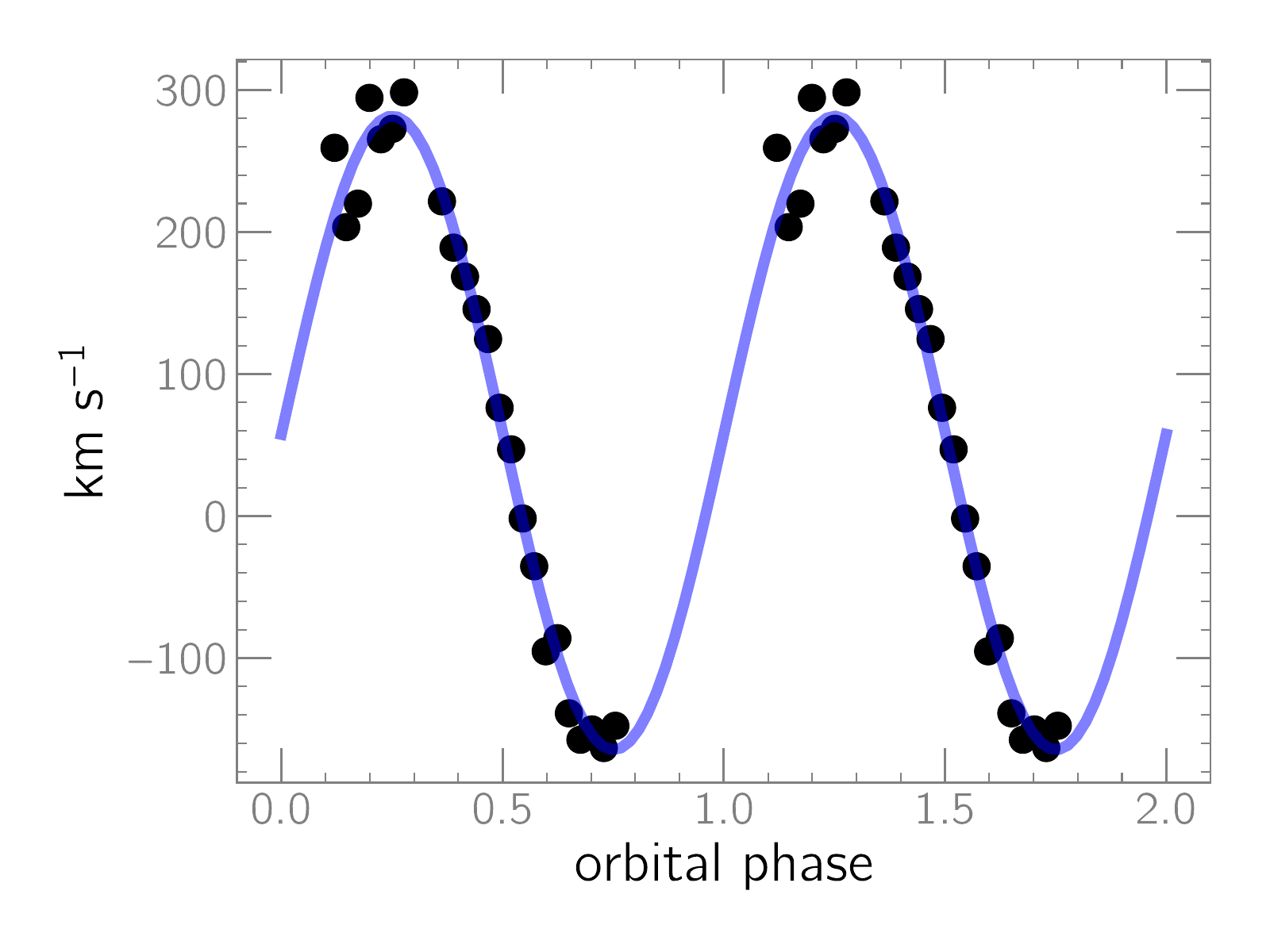}
        \caption{{\bf Top:} Continuum-subtracted, two-dimensional spectra of He~II~$\lambda 4686$~\AA\ and H$\beta$. There is no evidence of an eclipse at inferior conjunction. The two panels use different intensity cuts. {\bf Bottom:} Radial-velocity curve of the Ca~II triplet, showing a semiamplitude of 220~km~s$^{-1}$ with a systemic velocity of 60~km~s$^{-1}$. The blue-to-red crossing establishes the epoch of inferior conjunction. The Ca~II line fluxes near the time of inferior conjunction have been excluded because they are very low. This suggests that the secondary's inner hemisphere is occluded by the rest of the star near inferior conjunction, consistent with a moderate-to-high orbital inclination. 
        }
        \label{fig:LBT_spectra} 
    \end{figure}

    \subsection{Orbital ephemeris from LBT spectroscopy} \label{sec:ephem}

    The LBT spectra enable us to phase Paloma's \tess\ light curve to the binary orbit because several features from the donor star are present. In particular, there is significant emission at the Ca~II~$\lambda\lambda 8498, 8542, 8662$~\AA\ triplet from the secondary's inner hemisphere, and the Na~I~$\lambda\lambda 8183, 8195$~\AA\ absorption doublet is weakly present. In addition to the Ca~II triplet, Paloma's spectrum contains a large number of narrow, weak metal lines that are visible when the secondary's irradiated inner hemisphere is viewed preferentially (Fig.~\ref{fig:spectrum}).
    
    The Ca~II emission and the Na~I absorption move in phase with each other, and since the signal-to-noise ratio of the Ca~II lines is significantly higher, we measure their motion to obtain a radial velocity curve (Fig.~\ref{fig:LBT_spectra}). The blue-to-red crossing of the Ca~II lines yields the time of the donor's inferior conjunction, for which we provide an ephemeris of: \begin{equation} T_{conj}[BJD] = 2458836.9131(2) + 0.10914(12)  \times\ E. \label{ephem} \end{equation} Since the spectra were obtained during the \tess\ observation, the relative imprecision of the orbital period results in negligible phasing errors across the month-long \tess\ light curve.
    

    Given the complexity of the LBT spectra, it is beyond the scope of this paper to analyze them comprehensively, and we will do so in a separate paper. However, an initial analysis of the spectra does not offer a clear-cut explanation of the variable secondary spin maximum discussed earlier. The observations were obtained between beat phases 0.98-1.12, which is outside the interval during which the secondary maximum vanishes at inferior conjunction (Fig.~\ref{fig:paloma_2D_lightcurves}). The spectra do provide evidence of a moderately high orbital inclination, as evidenced by the precipitous decline of the secondary's metal emission lines hemisphere near inferior conjunction. This behavior, which is the reason for the lack of measured radial velocities for the Ca~II triplet near inferior conjunction in Fig.~\ref{fig:LBT_spectra}, is likely the result of the secondary's irradiated inner hemisphere being hidden by its backside. However, there is no evidence of an eclipse of the accretion flow in either the He~II or H$\beta$ lines in Fig.~\ref{fig:LBT_spectra}.

    \subsection{The spin period derivative of Paloma}
    
        With a sufficiently long ($\sim$decades) observational baseline, we would expect Paloma to show a spin-period derivative ($\dot{P}$), as has been observed in all APs with such baselines \citep[for a recent summary, see Table~1 in][]{myers}. These systems have a characteristic synchronization timescale $\tau = |(P_{orb} - P_{spin})| / \dot{P}$.
        
        We measure a spin period of 0.09460(10)~d at Julian year epoch 2019.94, compared to a period of 0.094622(3)~d in observations between 1992 and 2002 \citep{schwarz}. Because the \citet{schwarz} value falls within our 1$\sigma$ uncertainty for the spin period, we do not detect evidence of a statistically significant $\dot{P}$. This non-detection is subject to an important caveat: the \citet{schwarz} period suffers from a cycle-count ambiguity. 
        
        Although the non-detection of $\dot{P}$ is disappointing, it is not unexpected, given the relatively large uncertainty of the \tess\ spin period. The maximum observed $\dot{P}$ in FO~Aqr, an IP famous for its rapidly varying spin period, is $|\dot{P}| = 8\times10^{-10}$ \citep{littlefield20}. If the \citet{schwarz} period were either increasing or decreasing at that rate, the change in period would be indiscernible a quarter-century later at the precision of the \tess\ spin period. Without long-term, highly precise measurements of the spin period, it will be challenging to convincingly detect $\dot{P}$ in Paloma.

\section{Conclusion}

    We have used a long-cadence \ktwo\ light curve to show that J0846 is a new asynchronous polar with a significantly longer orbital period than any other AP. Our analysis of the \tess\ light curve of another nearly synchronous magnetic CV, Paloma, eliminates the long-standing ambiguity surrounding the proper identification of its spin frequency. Both targets warrant long-term monitoring so that their spin-period derivatives can be measured.

\acknowledgments

We thank the anonymous referee for their review of this manuscript.

PS, DH and CL acknowledge support from NSF grant AST-1514737. M.R.K acknowledges support from the Irish Research Council in the form of a Government of Ireland Postdoctoral Fellowship (GOIPD/2021/670: Invisible Monsters). PG and CL acknowledge support from NASA grants 80NSSC19K1704 and 80NSSC22K0183. KI acknowledges support from Polish National Science Center grant 2021/40/C/ST9/00186. P.A.M. acknowledges support from Picture Rocks Observatory. This research has made use of the International Variable Star Index (VSX) database, operated at AAVSO, Cambridge, Massachusetts, USA.

\facility{Large Binocular Telescope}
\software{ {\tt astropy} \citep{astropy}, {\tt lightkurve} \citep{lightkurve} }

\newpage

\bibliography{bib.bib}

\begin{thebibliography}{}
\expandafter\ifx\csname natexlab\endcsname\relax\def\natexlab#1{#1}\fi
\providecommand{\url}[1]{\href{#1}{#1}}
\providecommand{\dodoi}[1]{doi:~\href{http://doi.org/#1}{\nolinkurl{#1}}}
\providecommand{\doeprint}[1]{\href{http://ascl.net/#1}{\nolinkurl{http://ascl.net/#1}}}
\providecommand{\doarXiv}[1]{\href{https://arxiv.org/abs/#1}{\nolinkurl{https://arxiv.org/abs/#1}}}

\bibitem[{{Astropy Collaboration} {et~al.}(2013){Astropy Collaboration},
  {Robitaille}, {Tollerud}, {Greenfield}, {Droettboom}, {Bray}, {Aldcroft},
  {Davis}, {Ginsburg}, {Price-Whelan}, {Kerzendorf}, {Conley}, {Crighton},
  {Barbary}, {Muna}, {Ferguson}, {Grollier}, {Parikh}, {Nair}, {Unther},
  {Deil}, {Woillez}, {Conseil}, {Kramer}, {Turner}, {Singer}, {Fox}, {Weaver},
  {Zabalza}, {Edwards}, {Azalee Bostroem}, {Burke}, {Casey}, {Crawford},
  {Dencheva}, {Ely}, {Jenness}, {Labrie}, {Lim}, {Pierfederici}, {Pontzen},
  {Ptak}, {Refsdal}, {Servillat}, \& {Streicher}}]{astropy}
{Astropy Collaboration}, {Robitaille}, T.~P., {Tollerud}, E.~J., {et~al.} 2013,
  \aap, 558, A33, \dodoi{10.1051/0004-6361/201322068}

\bibitem[{{Bailer-Jones} {et~al.}(2021){Bailer-Jones}, {Rybizki}, {Fouesneau},
  {Demleitner}, \& {Andrae}}]{BJ20}
{Bailer-Jones}, C.~A.~L., {Rybizki}, J., {Fouesneau}, M., {Demleitner}, M., \&
  {Andrae}, R. 2021, \aj, 161, 147, \dodoi{10.3847/1538-3881/abd806}

\bibitem[{{Beardmore} {et~al.}(1998){Beardmore}, {Mukai}, {Norton}, {Osborne},
  \& {Hellier}}]{beardmore98}
{Beardmore}, A.~P., {Mukai}, K., {Norton}, A.~J., {Osborne}, J.~P., \&
  {Hellier}, C. 1998, \mnras, 297, 337,
  \dodoi{10.1046/j.1365-8711.1998.01382.x}

\bibitem[{{Bellm} {et~al.}(2019){Bellm}, {Kulkarni}, {Graham}, {Dekany},
  {Smith}, {Riddle}, {Masci}, {Helou}, {Prince}, {Adams}, {Barbarino},
  {Barlow}, {Bauer}, {Beck}, {Belicki}, {Biswas}, {Blagorodnova}, {Bodewits},
  {Bolin}, {Brinnel}, {Brooke}, {Bue}, {Bulla}, {Burruss}, {Cenko}, {Chang},
  {Connolly}, {Coughlin}, {Cromer}, {Cunningham}, {De}, {Delacroix}, {Desai},
  {Duev}, {Eadie}, {Farnham}, {Feeney}, {Feindt}, {Flynn}, {Franckowiak},
  {Frederick}, {Fremling}, {Gal-Yam}, {Gezari}, {Giomi}, {Goldstein},
  {Golkhou}, {Goobar}, {Groom}, {Hacopians}, {Hale}, {Henning}, {Ho}, {Hover},
  {Howell}, {Hung}, {Huppenkothen}, {Imel}, {Ip}, {Ivezi{\'c}}, {Jackson},
  {Jones}, {Juric}, {Kasliwal}, {Kaspi}, {Kaye}, {Kelley}, {Kowalski},
  {Kramer}, {Kupfer}, {Landry}, {Laher}, {Lee}, {Lin}, {Lin}, {Lunnan},
  {Giomi}, {Mahabal}, {Mao}, {Miller}, {Monkewitz}, {Murphy}, {Ngeow},
  {Nordin}, {Nugent}, {Ofek}, {Patterson}, {Penprase}, {Porter}, {Rauch},
  {Rebbapragada}, {Reiley}, {Rigault}, {Rodriguez}, {van Roestel}, {Rusholme},
  {van Santen}, {Schulze}, {Shupe}, {Singer}, {Soumagnac}, {Stein}, {Surace},
  {Sollerman}, {Szkody}, {Taddia}, {Terek}, {Van Sistine}, {van Velzen},
  {Vestrand}, {Walters}, {Ward}, {Ye}, {Yu}, {Yan}, \& {Zolkower}}]{bellm2019}
{Bellm}, E.~C., {Kulkarni}, S.~R., {Graham}, M.~J., {et~al.} 2019, \pasp, 131,
  018002, \dodoi{10.1088/1538-3873/aaecbe}

\bibitem[{{Beuermann} {et~al.}(2021){Beuermann}, {Burwitz}, {Reinsch},
  {Schwope}, \& {Thomas}}]{beuermann21}
{Beuermann}, K., {Burwitz}, V., {Reinsch}, K., {Schwope}, A., \& {Thomas},
  H.~C. 2021, \aap, 645, A56, \dodoi{10.1051/0004-6361/202038598}

\bibitem[{{Chanmugam} \& {Ray}(1984)}]{chanmugam}
{Chanmugam}, G., \& {Ray}, A. 1984, \apj, 285, 252, \dodoi{10.1086/162499}

\bibitem[{{Cropper}(1990)}]{cropper}
{Cropper}, M. 1990, \ssr, 54, 195, \dodoi{10.1007/BF00177799}

\bibitem[{{Drake} {et~al.}(2014){Drake}, {Graham}, {Djorgovski}, {Catelan},
  {Mahabal}, {Torrealba}, {Garc{\'\i}a-{\'A}lvarez}, {Donalek}, {Prieto},
  {Williams}, {Larson}, {Christen sen}, {Belokurov}, {Koposov}, {Beshore},
  {Boattini}, {Gibbs}, {Hill}, {Kowalski}, {Johnson}, \& {Shelly}}]{drake}
{Drake}, A.~J., {Graham}, M.~J., {Djorgovski}, S.~G., {et~al.} 2014, \apjs,
  213, 9, \dodoi{10.1088/0067-0049/213/1/9}

\bibitem[{{Esposito} {et~al.}(2015){Esposito}, {Israel}, {Milisavljevic},
  {Mapelli}, {Zampieri}, {Sidoli}, {Fabbiano}, \& {Rodr{\'\i}guez
  Castillo}}]{esposito}
{Esposito}, P., {Israel}, G.~L., {Milisavljevic}, D., {et~al.} 2015, \mnras,
  452, 1112, \dodoi{10.1093/mnras/stv1379}

\bibitem[{{Gaia Collaboration} {et~al.}(2016){Gaia Collaboration}, {Prusti},
  {de Bruijne}, {Brown}, {Vallenari}, {Babusiaux}, {Bailer-Jones}, {Bastian},
  {Biermann}, {Evans}, {Eyer}, {Jansen}, {Jordi}, {Klioner}, {Lammers},
  {Lindegren}, {Luri}, {Mignard}, {Milligan}, {Panem}, {Poinsignon},
  {Pourbaix}, {Randich}, {Sarri}, {Sartoretti}, {Siddiqui}, {Soubiran},
  {Valette}, {van Leeuwen}, {Walton}, {Aerts}, {Arenou}, {Cropper}, {Drimmel},
  {H{\o}g}, {Katz}, {Lattanzi}, {O'Mullane}, {Grebel}, {Holland}, {Huc},
  {Passot}, {Bramante}, {Cacciari}, {Casta{\~n}eda}, {Chaoul}, {Cheek}, {De
  Angeli}, {Fabricius}, {Guerra}, {Hern{\'a}ndez}, {Jean-Antoine-Piccolo},
  {Masana}, {Messineo}, {Mowlavi}, {Nienartowicz}, {Ord{\'o}{\~n}ez-Blanco},
  {Panuzzo}, {Portell}, {Richards}, {Riello}, {Seabroke}, {Tanga},
  {Th{\'e}venin}, {Torra}, {Els}, {Gracia-Abril}, {Comoretto},
  {Garcia-Reinaldos}, {Lock}, {Mercier}, {Altmann}, {Andrae}, {Astraatmadja},
  {Bellas-Velidis}, {Benson}, {Berthier}, {Blomme}, {Busso}, {Carry},
  {Cellino}, {Clementini}, {Cowell}, {Creevey}, {Cuypers}, {Davidson}, {De
  Ridder}, {de Torres}, {Delchambre}, {Dell'Oro}, {Ducourant}, {Fr{\'e}mat},
  {Garc{\'\i}a-Torres}, {Gosset}, {Halbwachs}, {Hambly}, {Harrison}, {Hauser},
  {Hestroffer}, {Hodgkin}, {Huckle}, {Hutton}, {Jasniewicz}, {Jordan},
  {Kontizas}, {Korn}, {Lanzafame}, {Manteiga}, {Moitinho}, {Muinonen},
  {Osinde}, {Pancino}, {Pauwels}, {Petit}, {Recio-Blanco}, {Robin}, {Sarro},
  {Siopis}, {Smith}, {Smith}, {Sozzetti}, {Thuillot}, {van Reeven}, {Viala},
  {Abbas}, {Abreu Aramburu}, {Accart}, {Aguado}, {Allan}, {Allasia},
  {Altavilla}, {{\'A}lvarez}, {Alves}, {Anderson}, {Andrei}, {Anglada Varela},
  {Antiche}, {Antoja}, {Ant{\'o}n}, {Arcay}, {Atzei}, {Ayache}, {Bach},
  {Baker}, {Balaguer-N{\'u}{\~n}ez}, {Barache}, {Barata}, {Barbier}, {Barblan},
  {Baroni}, {Barrado y Navascu{\'e}s}, {Barros}, {Barstow}, {Becciani},
  {Bellazzini}, {Bellei}, {Bello Garc{\'\i}a}, {Belokurov}, {Bendjoya},
  {Berihuete}, {Bianchi}, {Bienaym{\'e}}, {Billebaud}, {Blagorodnova},
  {Blanco-Cuaresma}, {Boch}, {Bombrun}, {Borrachero}, {Bouquillon}, {Bourda},
  {Bouy}, {Bragaglia}, {Breddels}, {Brouillet}, {Br{\"u}semeister},
  {Bucciarelli}, {Budnik}, {Burgess}, {Burgon}, {Burlacu}, {Busonero}, {Buzzi},
  {Caffau}, {Cambras}, {Campbell}, {Cancelliere}, {Cantat-Gaudin}, {Carlucci},
  {Carrasco}, {Castellani}, {Charlot}, {Charnas}, {Charvet}, {Chassat},
  {Chiavassa}, {Clotet}, {Cocozza}, {Collins}, {Collins}, {Costigan}, {Crifo},
  {Cross}, {Crosta}, {Crowley}, {Dafonte}, {Damerdji}, {Dapergolas}, {David},
  {David}, {De Cat}, {de Felice}, {de Laverny}, {De Luise}, {De March}, {de
  Martino}, {de Souza}, {Debosscher}, {del Pozo}, {Delbo}, {Delgado},
  {Delgado}, {di Marco}, {Di Matteo}, {Diakite}, {Distefano}, {Dolding}, {Dos
  Anjos}, {Drazinos}, {Dur{\'a}n}, {Dzigan}, {Ecale}, {Edvardsson}, {Enke},
  {Erdmann}, {Escolar}, {Espina}, {Evans}, {Eynard Bontemps}, {Fabre},
  {Fabrizio}, {Faigler}, {Falc{\~a}o}, {Farr{\`a}s Casas}, {Faye}, {Federici},
  {Fedorets}, {Fern{\'a}ndez-Hern{\'a}ndez}, {Fernique}, {Fienga}, {Figueras},
  {Filippi}, {Findeisen}, {Fonti}, {Fouesneau}, {Fraile}, {Fraser}, {Fuchs},
  {Furnell}, {Gai}, {Galleti}, {Galluccio}, {Garabato}, {Garc{\'\i}a-Sedano},
  {Gar{\'e}}, {Garofalo}, {Garralda}, {Gavras}, {Gerssen}, {Geyer}, {Gilmore},
  {Girona}, {Giuffrida}, {Gomes}, {Gonz{\'a}lez-Marcos},
  {Gonz{\'a}lez-N{\'u}{\~n}ez}, {Gonz{\'a}lez-Vidal}, {Granvik}, {Guerrier},
  {Guillout}, {Guiraud}, {G{\'u}rpide}, {Guti{\'e}rrez-S{\'a}nchez}, {Guy},
  {Haigron}, {Hatzidimitriou}, {Haywood}, {Heiter}, {Helmi}, {Hobbs},
  {Hofmann}, {Holl}, {Holland}, {Hunt}, {Hypki}, {Icardi}, {Irwin}, {Jevardat
  de Fombelle}, {Jofr{\'e}}, {Jonker}, {Jorissen}, {Julbe}, {Karampelas},
  {Kochoska}, {Kohley}, {Kolenberg}, {Kontizas}, {Koposov}, {Kordopatis},
  {Koubsky}, {Kowalczyk}, {Krone-Martins}, {Kudryashova}, {Kull}, {Bachchan},
  {Lacoste-Seris}, {Lanza}, {Lavigne}, {Le Poncin-Lafitte}, {Lebreton},
  {Lebzelter}, {Leccia}, {Leclerc}, {Lecoeur-Taibi}, {Lemaitre}, {Lenhardt},
  {Leroux}, {Liao}, {Licata}, {Lindstr{\o}m}, {Lister}, {Livanou}, {Lobel},
  {L{\"o}ffler}, {L{\'o}pez}, {Lopez-Lozano}, {Lorenz}, {Loureiro},
  {MacDonald}, {Magalh{\~a}es Fernandes}, {Managau}, {Mann}, {Mantelet},
  {Marchal}, {Marchant}, {Marconi}, {Marie}, {Marinoni}, {Marrese},
  {Marschalk{\'o}}, {Marshall}, {Mart{\'\i}n-Fleitas}, {Martino}, {Mary},
  {Matijevi{\v{c}}}, {Mazeh}, {McMillan}, {Messina}, {Mestre}, {Michalik},
  {Millar}, {Miranda}, {Molina}, {Molinaro}, {Molinaro}, {Moln{\'a}r},
  {Moniez}, {Montegriffo}, {Monteiro}, {Mor}, {Mora}, {Morbidelli}, {Morel},
  {Morgenthaler}, {Morley}, {Morris}, {Mulone}, {Muraveva}, {Musella},
  {Narbonne}, {Nelemans}, {Nicastro}, {Noval}, {Ord{\'e}novic},
  {Ordieres-Mer{\'e}}, {Osborne}, {Pagani}, {Pagano}, {Pailler}, {Palacin},
  {Palaversa}, {Parsons}, {Paulsen}, {Pecoraro}, {Pedrosa}, {Pentik{\"a}inen},
  {Pereira}, {Pichon}, {Piersimoni}, {Pineau}, {Plachy}, {Plum}, {Poujoulet},
  {Pr{\v{s}}a}, {Pulone}, {Ragaini}, {Rago}, {Rambaux}, {Ramos-Lerate},
  {Ranalli}, {Rauw}, {Read}, {Regibo}, {Renk}, {Reyl{\'e}}, {Ribeiro},
  {Rimoldini}, {Ripepi}, {Riva}, {Rixon}, {Roelens}, {Romero-G{\'o}mez},
  {Rowell}, {Royer}, {Rudolph}, {Ruiz-Dern}, {Sadowski}, {Sagrist{\`a}
  Sell{\'e}s}, {Sahlmann}, {Salgado}, {Salguero}, {Sarasso}, {Savietto},
  {Schnorhk}, {Schultheis}, {Sciacca}, {Segol}, {Segovia}, {Segransan},
  {Serpell}, {Shih}, {Smareglia}, {Smart}, {Smith}, {Solano}, {Solitro},
  {Sordo}, {Soria Nieto}, {Souchay}, {Spagna}, {Spoto}, {Stampa}, {Steele},
  {Steidelm{\"u}ller}, {Stephenson}, {Stoev}, {Suess}, {S{\"u}veges}, {Surdej},
  {Szabados}, {Szegedi-Elek}, {Tapiador}, {Taris}, {Tauran}, {Taylor},
  {Teixeira}, {Terrett}, {Tingley}, {Trager}, {Turon}, {Ulla}, {Utrilla},
  {Valentini}, {van Elteren}, {Van Hemelryck}, {van Leeuwen}, {Varadi},
  {Vecchiato}, {Veljanoski}, {Via}, {Vicente}, {Vogt}, {Voss}, {Votruba},
  {Voutsinas}, {Walmsley}, {Weiler}, {Weingrill}, {Werner}, {Wevers},
  {Whitehead}, {Wyrzykowski}, {Yoldas}, {{\v{Z}}erjal}, {Zucker}, {Zurbach},
  {Zwitter}, {Alecu}, {Allen}, {Allende Prieto}, {Amorim},
  {Anglada-Escud{\'e}}, {Arsenijevic}, {Azaz}, {Balm}, {Beck}, {Bernstein},
  {Bigot}, {Bijaoui}, {Blasco}, {Bonfigli}, {Bono}, {Boudreault}, {Bressan},
  {Brown}, {Brunet}, {Bunclark}, {Buonanno}, {Butkevich}, {Carret}, {Carrion},
  {Chemin}, {Ch{\'e}reau}, {Corcione}, {Darmigny}, {de Boer}, {de Teodoro}, {de
  Zeeuw}, {Delle Luche}, {Domingues}, {Dubath}, {Fodor}, {Fr{\'e}zouls},
  {Fries}, {Fustes}, {Fyfe}, {Gallardo}, {Gallegos}, {Gardiol}, {Gebran},
  {Gomboc}, {G{\'o}mez}, {Grux}, {Gueguen}, {Heyrovsky}, {Hoar}, {Iannicola},
  {Isasi Parache}, {Janotto}, {Joliet}, {Jonckheere}, {Keil}, {Kim},
  {Klagyivik}, {Klar}, {Knude}, {Kochukhov}, {Kolka}, {Kos}, {Kutka}, {Lainey},
  {LeBouquin}, {Liu}, {Loreggia}, {Makarov}, {Marseille}, {Martayan},
  {Martinez-Rubi}, {Massart}, {Meynadier}, {Mignot}, {Munari}, {Nguyen},
  {Nordlander}, {Ocvirk}, {O'Flaherty}, {Olias Sanz}, {Ortiz}, {Osorio},
  {Oszkiewicz}, {Ouzounis}, {Palmer}, {Park}, {Pasquato}, {Peltzer}, {Peralta},
  {P{\'e}turaud}, {Pieniluoma}, {Pigozzi}, {Poels}, {Prat}, {Prod'homme},
  {Raison}, {Rebordao}, {Risquez}, {Rocca-Volmerange}, {Rosen}, {Ruiz-Fuertes},
  {Russo}, {Sembay}, {Serraller Vizcaino}, {Short}, {Siebert}, {Silva},
  {Sinachopoulos}, {Slezak}, {Soffel}, {Sosnowska}, {Strai{\v{z}}ys}, {ter
  Linden}, {Terrell}, {Theil}, {Tiede}, {Troisi}, {Tsalmantza}, {Tur},
  {Vaccari}, {Vachier}, {Valles}, {Van Hamme}, {Veltz}, {Virtanen}, {Wallut},
  {Wichmann}, {Wilkinson}, {Ziaeepour}, \& {Zschocke}}]{gaia}
{Gaia Collaboration}, {Prusti}, T., {de Bruijne}, J.~H.~J., {et~al.} 2016,
  \aap, 595, A1, \dodoi{10.1051/0004-6361/201629272}

\bibitem[{{Gaia Collaboration} {et~al.}(2021){Gaia Collaboration}, {Brown},
  {Vallenari}, {Prusti}, {de Bruijne}, {Babusiaux}, {Biermann}, {Creevey},
  {Evans}, {Eyer}, {Hutton}, {Jansen}, {Jordi}, {Klioner}, {Lammers},
  {Lindegren}, {Luri}, {Mignard}, {Panem}, {Pourbaix}, {Randich}, {Sartoretti},
  {Soubiran}, {Walton}, {Arenou}, {Bailer-Jones}, {Bastian}, {Cropper},
  {Drimmel}, {Katz}, {Lattanzi}, {van Leeuwen}, {Bakker}, {Cacciari},
  {Casta{\~n}eda}, {De Angeli}, {Ducourant}, {Fabricius}, {Fouesneau},
  {Fr{\'e}mat}, {Guerra}, {Guerrier}, {Guiraud}, {Jean-Antoine Piccolo},
  {Masana}, {Messineo}, {Mowlavi}, {Nicolas}, {Nienartowicz}, {Pailler},
  {Panuzzo}, {Riclet}, {Roux}, {Seabroke}, {Sordo}, {Tanga}, {Th{\'e}venin},
  {Gracia-Abril}, {Portell}, {Teyssier}, {Altmann}, {Andrae}, {Bellas-Velidis},
  {Benson}, {Berthier}, {Blomme}, {Brugaletta}, {Burgess}, {Busso}, {Carry},
  {Cellino}, {Cheek}, {Clementini}, {Damerdji}, {Davidson}, {Delchambre},
  {Dell'Oro}, {Fern{\'a}ndez-Hern{\'a}ndez}, {Galluccio}, {Garc{\'\i}a-Lario},
  {Garcia-Reinaldos}, {Gonz{\'a}lez-N{\'u}{\~n}ez}, {Gosset}, {Haigron},
  {Halbwachs}, {Hambly}, {Harrison}, {Hatzidimitriou}, {Heiter},
  {Hern{\'a}ndez}, {Hestroffer}, {Hodgkin}, {Holl}, {Jan{\ss}en}, {Jevardat de
  Fombelle}, {Jordan}, {Krone-Martins}, {Lanzafame}, {L{\"o}ffler}, {Lorca},
  {Manteiga}, {Marchal}, {Marrese}, {Moitinho}, {Mora}, {Muinonen}, {Osborne},
  {Pancino}, {Pauwels}, {Petit}, {Recio-Blanco}, {Richards}, {Riello},
  {Rimoldini}, {Robin}, {Roegiers}, {Rybizki}, {Sarro}, {Siopis}, {Smith},
  {Sozzetti}, {Ulla}, {Utrilla}, {van Leeuwen}, {van Reeven}, {Abbas}, {Abreu
  Aramburu}, {Accart}, {Aerts}, {Aguado}, {Ajaj}, {Altavilla}, {{\'A}lvarez},
  {{\'A}lvarez Cid-Fuentes}, {Alves}, {Anderson}, {Anglada Varela}, {Antoja},
  {Audard}, {Baines}, {Baker}, {Balaguer-N{\'u}{\~n}ez}, {Balbinot}, {Balog},
  {Barache}, {Barbato}, {Barros}, {Barstow}, {Bartolom{\'e}}, {Bassilana},
  {Bauchet}, {Baudesson-Stella}, {Becciani}, {Bellazzini}, {Bernet}, {Bertone},
  {Bianchi}, {Blanco-Cuaresma}, {Boch}, {Bombrun}, {Bossini}, {Bouquillon},
  {Bragaglia}, {Bramante}, {Breedt}, {Bressan}, {Brouillet}, {Bucciarelli},
  {Burlacu}, {Busonero}, {Butkevich}, {Buzzi}, {Caffau}, {Cancelliere},
  {C{\'a}novas}, {Cantat-Gaudin}, {Carballo}, {Carlucci}, {Carnerero},
  {Carrasco}, {Casamiquela}, {Castellani}, {Castro-Ginard}, {Castro Sampol},
  {Chaoul}, {Charlot}, {Chemin}, {Chiavassa}, {Cioni}, {Comoretto}, {Cooper},
  {Cornez}, {Cowell}, {Crifo}, {Crosta}, {Crowley}, {Dafonte}, {Dapergolas},
  {David}, {David}, {de Laverny}, {De Luise}, {De March}, {De Ridder}, {de
  Souza}, {de Teodoro}, {de Torres}, {del Peloso}, {del Pozo}, {Delbo},
  {Delgado}, {Delgado}, {Delisle}, {Di Matteo}, {Diakite}, {Diener},
  {Distefano}, {Dolding}, {Eappachen}, {Edvardsson}, {Enke}, {Esquej}, {Fabre},
  {Fabrizio}, {Faigler}, {Fedorets}, {Fernique}, {Fienga}, {Figueras},
  {Fouron}, {Fragkoudi}, {Fraile}, {Franke}, {Gai}, {Garabato},
  {Garcia-Gutierrez}, {Garc{\'\i}a-Torres}, {Garofalo}, {Gavras}, {Gerlach},
  {Geyer}, {Giacobbe}, {Gilmore}, {Girona}, {Giuffrida}, {Gomel}, {Gomez},
  {Gonzalez-Santamaria}, {Gonz{\'a}lez-Vidal}, {Granvik},
  {Guti{\'e}rrez-S{\'a}nchez}, {Guy}, {Hauser}, {Haywood}, {Helmi}, {Hidalgo},
  {Hilger}, {H{\l}adczuk}, {Hobbs}, {Holland}, {Huckle}, {Jasniewicz},
  {Jonker}, {Juaristi Campillo}, {Julbe}, {Karbevska}, {Kervella}, {Khanna},
  {Kochoska}, {Kontizas}, {Kordopatis}, {Korn}, {Kostrzewa-Rutkowska},
  {Kruszy{\'n}ska}, {Lambert}, {Lanza}, {Lasne}, {Le Campion}, {Le Fustec},
  {Lebreton}, {Lebzelter}, {Leccia}, {Leclerc}, {Lecoeur-Taibi}, {Liao},
  {Licata}, {Lindstr{\o}m}, {Lister}, {Livanou}, {Lobel}, {Madrero Pardo},
  {Managau}, {Mann}, {Marchant}, {Marconi}, {Marcos Santos}, {Marinoni},
  {Marocco}, {Marshall}, {Martin Polo}, {Mart{\'\i}n-Fleitas}, {Masip},
  {Massari}, {Mastrobuono-Battisti}, {Mazeh}, {McMillan}, {Messina},
  {Michalik}, {Millar}, {Mints}, {Molina}, {Molinaro}, {Moln{\'a}r},
  {Montegriffo}, {Mor}, {Morbidelli}, {Morel}, {Morris}, {Mulone}, {Munoz},
  {Muraveva}, {Murphy}, {Musella}, {Noval}, {Ord{\'e}novic}, {Orr{\`u}},
  {Osinde}, {Pagani}, {Pagano}, {Palaversa}, {Palicio}, {Panahi}, {Pawlak},
  {Pe{\~n}alosa Esteller}, {Penttil{\"a}}, {Piersimoni}, {Pineau}, {Plachy},
  {Plum}, {Poggio}, {Poretti}, {Poujoulet}, {Pr{\v{s}}a}, {Pulone}, {Racero},
  {Ragaini}, {Rainer}, {Raiteri}, {Rambaux}, {Ramos}, {Ramos-Lerate}, {Re
  Fiorentin}, {Regibo}, {Reyl{\'e}}, {Ripepi}, {Riva}, {Rixon}, {Robichon},
  {Robin}, {Roelens}, {Rohrbasser}, {Romero-G{\'o}mez}, {Rowell}, {Royer},
  {Rybicki}, {Sadowski}, {Sagrist{\`a} Sell{\'e}s}, {Sahlmann}, {Salgado},
  {Salguero}, {Samaras}, {Sanchez Gimenez}, {Sanna}, {Santove{\~n}a},
  {Sarasso}, {Schultheis}, {Sciacca}, {Segol}, {Segovia}, {S{\'e}gransan},
  {Semeux}, {Shahaf}, {Siddiqui}, {Siebert}, {Siltala}, {Slezak}, {Smart},
  {Solano}, {Solitro}, {Souami}, {Souchay}, {Spagna}, {Spoto}, {Steele},
  {Steidelm{\"u}ller}, {Stephenson}, {S{\"u}veges}, {Szabados}, {Szegedi-Elek},
  {Taris}, {Tauran}, {Taylor}, {Teixeira}, {Thuillot}, {Tonello}, {Torra},
  {Torra}, {Turon}, {Unger}, {Vaillant}, {van Dillen}, {Vanel}, {Vecchiato},
  {Viala}, {Vicente}, {Voutsinas}, {Weiler}, {Wevers}, {Wyrzykowski}, {Yoldas},
  {Yvard}, {Zhao}, {Zorec}, {Zucker}, {Zurbach}, \& {Zwitter}}]{edr3}
{Gaia Collaboration}, {Brown}, A.~G.~A., {Vallenari}, A., {et~al.} 2021, \aap,
  649, A1, \dodoi{10.1051/0004-6361/202039657}

\bibitem[{{Geckeler} \& {Staubert}(1997)}]{gs97}
{Geckeler}, R.~D., \& {Staubert}, R. 1997, \aap, 325, 1070

\bibitem[{{Hakala} {et~al.}(2019){Hakala}, {Ramsay}, {Potter}, {Beardmore},
  {Buckley}, \& {Wynn}}]{hakala}
{Hakala}, P., {Ramsay}, G., {Potter}, S.~B., {et~al.} 2019, \mnras, 486, 2549,
  \dodoi{10.1093/mnras/stz992}

\bibitem[{{Halpern} {et~al.}(2017){Halpern}, {Bogdanov}, \&
  {Thorstensen}}]{halpern}
{Halpern}, J.~P., {Bogdanov}, S., \& {Thorstensen}, J.~R. 2017, \apj, 838, 124,
  \dodoi{10.3847/1538-4357/838/2/124}

\bibitem[{{Hill} {et~al.}(2022){Hill}, {Littlefield}, {Garnavich}, {Scaringi},
  {Szkody}, {Mason}, {Kennedy}, {Shaw}, \& {Covington}}]{hill22}
{Hill}, K.~L., {Littlefield}, C., {Garnavich}, P., {et~al.} 2022, arXiv
  e-prints, arXiv:2203.00221.
\newblock \doarXiv{2203.00221}

\bibitem[{{Joshi} {et~al.}(2016){Joshi}, {Pandey}, {Singh}, \&
  {Agrawal}}]{joshi}
{Joshi}, A., {Pandey}, J.~C., {Singh}, K.~P., \& {Agrawal}, P.~C. 2016, \apj,
  830, 56, \dodoi{10.3847/0004-637X/830/2/56}

\bibitem[{{King}(1993)}]{king}
{King}, A.~R. 1993, \mnras, 261, 144, \dodoi{10.1093/mnras/261.1.144}

\bibitem[{{King} \& {Lasota}(1991)}]{kl91}
{King}, A.~R., \& {Lasota}, J.-P. 1991, \apj, 378, 674, \dodoi{10.1086/170467}

\bibitem[{{King} \& {Wynn}(1999)}]{kw99}
{King}, A.~R., \& {Wynn}, G.~A. 1999, \mnras, 310, 203,
  \dodoi{10.1046/j.1365-8711.1999.02974.x}

\bibitem[{{Lightkurve Collaboration} {et~al.}(2018){Lightkurve Collaboration},
  {Cardoso}, {Hedges}, {Gully-Santiago}, {Saunders}, {Cody}, {Barclay}, {Hall},
  {Sagear}, {Turtelboom}, {Zhang}, {Tzanidakis}, {Mighell}, {Coughlin}, {Bell},
  {Berta-Thompson}, {Williams}, {Dotson}, \& {Barentsen}}]{lightkurve}
{Lightkurve Collaboration}, {Cardoso}, J. V. d.~M., {Hedges}, C., {et~al.}
  2018, {Lightkurve: Kepler and TESS time series analysis in Python}.
\newblock \doeprint{1812.013}

\bibitem[{{Littlefield} {et~al.}(2019){Littlefield}, {Garnavich}, {Mukai},
  {Mason}, {Szkody}, {Kennedy}, {Myers}, \& {Schwarz}}]{littlefield}
{Littlefield}, C., {Garnavich}, P., {Mukai}, K., {et~al.} 2019, \apj, 881, 141,
  \dodoi{10.3847/1538-4357/ab2a17}

\bibitem[{{Littlefield} {et~al.}(2020){Littlefield}, {Garnavich}, {Szkody},
  {Ramsay}, {Howell}, {Lima}, {Kennedy}, \& {Cook}}]{littlefield20}
{Littlefield}, C., {Garnavich}, P., {Szkody}, P., {et~al.} 2020, arXiv
  e-prints, arXiv:2004.08923.
\newblock \doarXiv{2004.08923}

\bibitem[{{Littlefield} {et~al.}(2021){Littlefield}, {Scaringi}, {Garnavich},
  {Szkody}, {Kennedy}, {I{\l}kiewicz}, \& {Mason}}]{littlefield21}
{Littlefield}, C., {Scaringi}, S., {Garnavich}, P., {et~al.} 2021, \aj, 162,
  49, \dodoi{10.3847/1538-3881/ac062b}

\bibitem[{{Littlefield} {et~al.}(2015){Littlefield}, {Mukai}, {Mumme}, {Cain},
  {Magno}, {Corpuz}, {Sand efur}, {Boyd}, {Cook}, {Ulowetz}, \&
  {Martinez}}]{littlefield15}
{Littlefield}, C., {Mukai}, K., {Mumme}, R., {et~al.} 2015, \mnras, 449, 3107,
  \dodoi{10.1093/mnras/stv462}

\bibitem[{{Lomb}(1976)}]{lomb1976}
{Lomb}, N.~R. 1976, \apss, 39, 447, \dodoi{10.1007/BF00648343}

\bibitem[{{Mason} {et~al.}(1995){Mason}, {Andronov}, {Kolesnikov}, {Pavlenko},
  \& {Shakovskoy}}]{mason95}
{Mason}, P.~A., {Andronov}, I.~L., {Kolesnikov}, S.~V., {Pavlenko}, E.~P., \&
  {Shakovskoy}, M. 1995, in Astronomical Society of the Pacific Conference
  Series, Vol.~85, Magnetic Cataclysmic Variables, ed. D.~A.~H. {Buckley} \&
  B.~{Warner}, 496

\bibitem[{{Mason} {et~al.}(1989){Mason}, {Liebert}, \& {Schmidt}}]{mason89}
{Mason}, P.~A., {Liebert}, J., \& {Schmidt}, G.~D. 1989, \apj, 346, 941,
  \dodoi{10.1086/168074}

\bibitem[{{Mason} {et~al.}(1998){Mason}, {Ramsay}, {Andronov}, {Kolesnikov},
  {Shakhovskoy}, \& {Pavlenko}}]{mason98}
{Mason}, P.~A., {Ramsay}, G., {Andronov}, I., {et~al.} 1998, \mnras, 295, 511,
  \dodoi{10.1046/j.1365-8711.1998.01185.x}

\bibitem[{{Mason} {et~al.}(2020){Mason}, {Morales}, {Littlefield}, {Garnavich},
  {Pavlenko}, {Szkody}, {Kennedy}, {Myers}, {Schwarz}, {Babina}, {Sosnovskij},
  {Antonyuk}, {Shugarov}, \& {Andreev}}]{mason20}
{Mason}, P.~A., {Morales}, J.~F., {Littlefield}, C., {et~al.} 2020, Advances in
  Space Research, 66, 1123, \dodoi{10.1016/j.asr.2020.03.038}

\bibitem[{{Myers} {et~al.}(2017){Myers}, {Patterson}, {de Miguel}, {Hambsch},
  {Monard}, {Bolt}, {McCormick}, {Rea}, \& {Allen}}]{myers}
{Myers}, G., {Patterson}, J., {de Miguel}, E., {et~al.} 2017, \pasp, 129,
  044204, \dodoi{10.1088/1538-3873/aa54a8}

\bibitem[{{Patterson}(1994)}]{patterson}
{Patterson}, J. 1994, \pasp, 106, 209, \dodoi{10.1086/133375}

\bibitem[{{Pavlenko} {et~al.}(2013){Pavlenko}, {Andreev}, {Babina}, \&
  {Malanushenko}}]{pavlenko_bycam}
{Pavlenko}, E., {Andreev}, M., {Babina}, Y., \& {Malanushenko}, V. 2013, in
  Astronomical Society of the Pacific Conference Series, Vol. 469, 18th
  European White Dwarf Workshop., ed. J.~{Krzesi{\'n}ski}, G.~{Stachowski},
  P.~{Moskalik}, \& K.~{Bajan}, 343

\bibitem[{{Pavlenko} {et~al.}(2018){Pavlenko}, {Mason}, {Sosnovskij},
  {Shugarov}, {Babina}, {Antonyuk}, {Andreev}, {Pit}, {Antonyuk}, \&
  {Baklanov}}]{pavlenko_v1500cyg}
{Pavlenko}, E.~P., {Mason}, P.~A., {Sosnovskij}, A.~A., {et~al.} 2018, \mnras,
  479, 341, \dodoi{10.1093/mnras/sty1494}

\bibitem[{{Pogge} {et~al.}(2010){Pogge}, {Atwood}, {Brewer}, {Byard},
  {Derwent}, {Gonzalez}, {Martini}, {Mason}, {O'Brien}, {Osmer}, {Pappalardo},
  {Steinbrecher}, {Teiga}, \& {Zhelem}}]{PoggeMODS}
{Pogge}, R.~W., {Atwood}, B., {Brewer}, D.~F., {et~al.} 2010, in Society of
  Photo-Optical Instrumentation Engineers (SPIE) Conference Series, Vol. 7735,
  77350A

\bibitem[{{Qiu} {et~al.}(2019){Qiu}, {Soria}, {Wang}, {Wiktorowicz}, {Liu},
  {Bai}, {Bogomazov}, {Di Stefano}, {Walton}, \& {Xu}}]{qiu}
{Qiu}, Y., {Soria}, R., {Wang}, S., {et~al.} 2019, \apj, 877, 57,
  \dodoi{10.3847/1538-4357/ab16e7}

\bibitem[{{Rawat} {et~al.}(2021){Rawat}, {Pandey}, \& {Joshi}}]{rawat}
{Rawat}, N., {Pandey}, J.~C., \& {Joshi}, A. 2021, \apj, 912, 78,
  \dodoi{10.3847/1538-4357/abedae}

\bibitem[{{Scargle}(1982)}]{scargle1982}
{Scargle}, J.~D. 1982, \apj, 263, 835, \dodoi{10.1086/160554}

\bibitem[{{Schmidt} \& {Stockman}(1991)}]{ss91}
{Schmidt}, G.~D., \& {Stockman}, H.~S. 1991, \apj, 371, 749,
  \dodoi{10.1086/169939}

\bibitem[{{Schwarz} {et~al.}(2007){Schwarz}, {Schwope}, {Staude}, {Rau},
  {Hasinger}, {Urrutia}, \& {Motch}}]{schwarz}
{Schwarz}, R., {Schwope}, A.~D., {Staude}, A., {et~al.} 2007, \aap, 473, 511,
  \dodoi{10.1051/0004-6361:20077684}

\bibitem[{{Stockman} {et~al.}(1988){Stockman}, {Schmidt}, \& {Lamb}}]{ssl88}
{Stockman}, H.~S., {Schmidt}, G.~D., \& {Lamb}, D.~Q. 1988, \apj, 332, 282,
  \dodoi{10.1086/166652}

\bibitem[{{Szkody} {et~al.}(2006){Szkody}, {Henden}, {Ag{\"u}eros}, {Anderson},
  {Bochanski}, {Knapp}, {Mannikko}, {Mukadam}, {Silvestri}, {Schmidt},
  {Stephanik}, {Watson}, {West}, {Winget}, {Wolfe}, {Barentine}, {Brinkmann},
  {Brewington}, {Downes}, {Harvanek}, {Kleinman}, {Krzesinski}, {Long},
  {Neilsen}, {Nitta}, {Schneider}, {Snedden}, \& {Voges}}]{szkody06}
{Szkody}, P., {Henden}, A., {Ag{\"u}eros}, M., {et~al.} 2006, \aj, 131, 973,
  \dodoi{10.1086/499308}

\bibitem[{{Tovmassian} {et~al.}(2017){Tovmassian}, {Gonz{\'a}lez-Buitrago},
  {Thorstensen}, {Kotze}, {Breytenbach}, {Schwope}, {Bernardini}, {Zharikov},
  {Hernandez}, {Buckley}, {de Miguel}, {Hambsch}, {Myers}, {Goff}, {Cejudo},
  {Starkey}, {Campbell}, {Ulowetz}, {Stein}, {Nelson}, {Reichart}, {Haislip},
  {Ivarsen}, {LaCluyze}, {Moore}, \& {Miroshnichenko}}]{tovmassian}
{Tovmassian}, G., {Gonz{\'a}lez-Buitrago}, D., {Thorstensen}, J., {et~al.}
  2017, \aap, 608, A36, \dodoi{10.1051/0004-6361/201731323}

\bibitem[{{Wang} {et~al.}(2020){Wang}, {Qian}, {Han}, {Fang}, {Zang}, \&
  {Liu}}]{wang2020}
{Wang}, Q., {Qian}, S., {Han}, Z., {et~al.} 2020, \apj, 892, 38,
  \dodoi{10.3847/1538-4357/ab7759}

\bibitem[{{Wynn} \& {King}(1992)}]{wk92}
{Wynn}, G.~A., \& {King}, A.~R. 1992, \mnras, 255, 83,
  \dodoi{10.1093/mnras/255.1.83}

\bibitem[{{Zhilkin} {et~al.}(2012){Zhilkin}, {Bisikalo}, \&
  {Mason}}]{zhilkin12}
{Zhilkin}, A.~G., {Bisikalo}, D.~V., \& {Mason}, P.~A. 2012, Astronomy Reports,
  56, 257, \dodoi{10.1134/S1063772912040087}

\bibitem[{{Zhilkin} {et~al.}(2016){Zhilkin}, {Bisikalo}, \&
  {Mason}}]{zhilkin16}
{Zhilkin}, A.~G., {Bisikalo}, D.~V., \& {Mason}, P.~A. 2016, in American
  Institute of Physics Conference Series, Vol. 1714, Space Plasma Physics,
  020002, \dodoi{10.1063/1.4942564}

\end{thebibliography}

\appendix

\section{Beat-phase resolved spin profiles of J0846 and Paloma}

    The \ktwo\ and \tess\ data showcase the gradual evolution of the spin profiles of both J0846 and Paloma across their respective beat cycles. Figs.~\ref{fig:2D_lightcurves}~and~\ref{fig:paloma_2D_lightcurves} used two-dimensional light curves to illustrate this behavior. 
    
\begin{figure*}
                \centering
                \includegraphics[width=\textwidth]{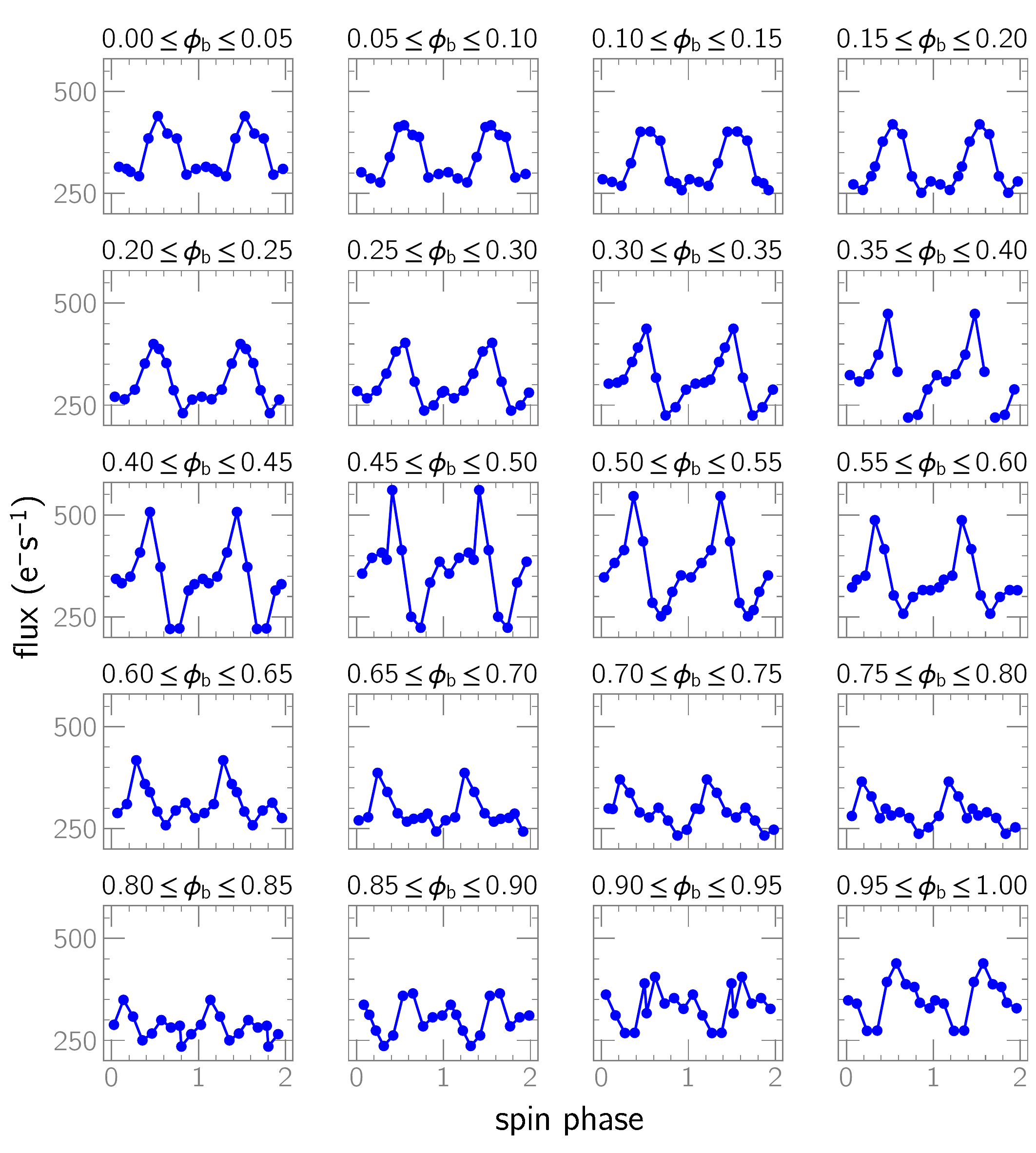}
                \caption{Binned spin profiles of J0846 in twenty non-overlapping portions of the 6.7~d beat cycle. The top of each panel indicates which beat phases were used to construct each spin profile. The data are repeated along the $x$ axis for clarity. The pole near spin phase 0.5 has an associated photometric maximum for most of the beat cycle, but it becomes indistinct near beat phase 0.8. 
                }
                \label{fig:J0846_spin_profiles}
        \end{figure*}

\begin{figure*}
        \centering
        \includegraphics[width=\textwidth]{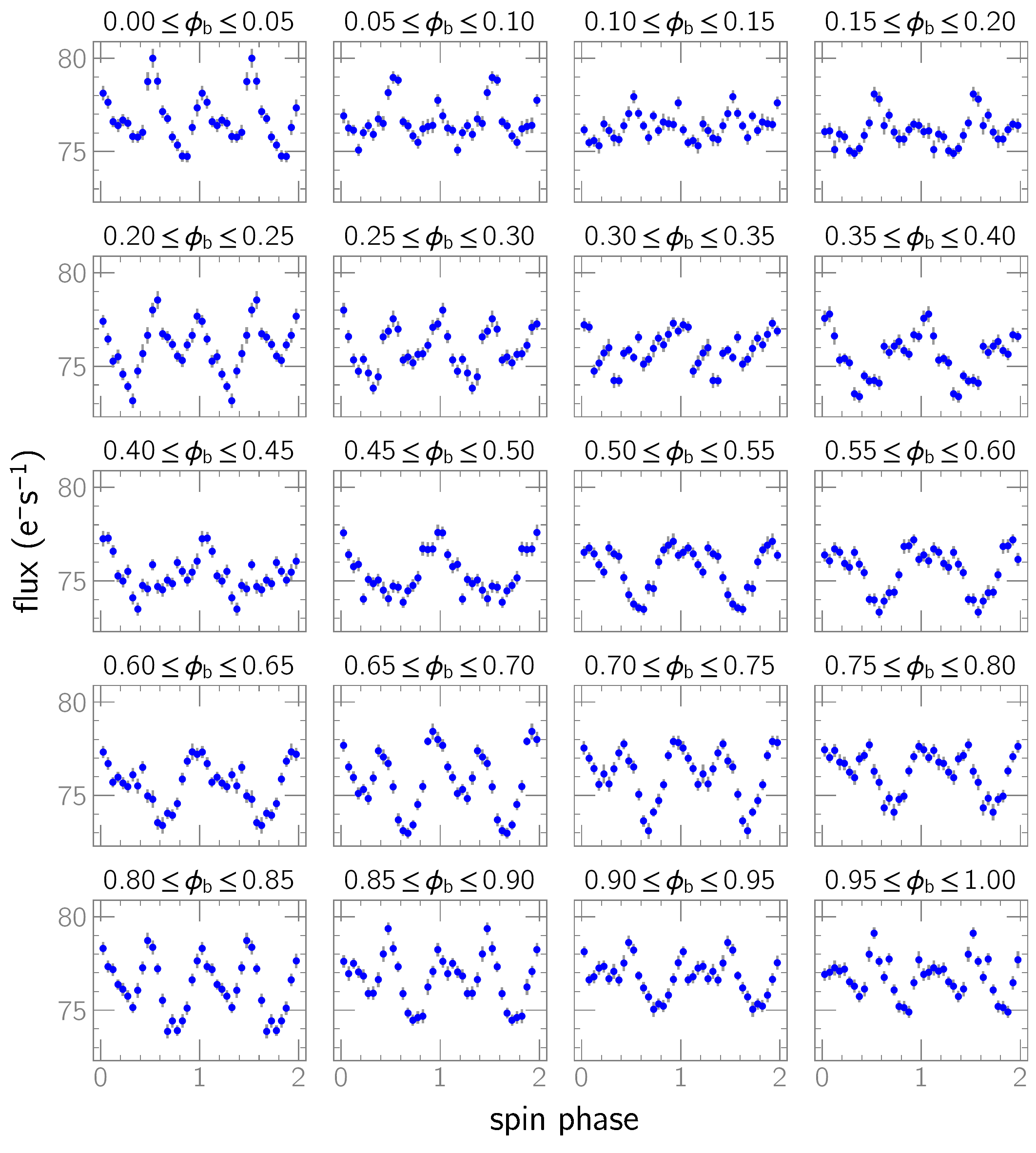}
        \caption{Evolution of Paloma's spin profile across the beat cycle. Beat phase 0.0 occurs when the primary spin maximum coincides with inferior conjunction of the secondary star (\textit{i.e.}, when the spin and orbital phases are both 0.0). }
        \label{fig:paloma_spin_profiles}
        
    \end{figure*}

    Here, we present one-dimensional light curves of the spin profiles of both J0846 and Paloma to enable a more careful inspection than is possible in their two-dimensional counterparts. Because the resulting figures are awkwardly large, we present them separately from the main text in Figs.~\ref{fig:J0846_spin_profiles}~and~\ref{fig:paloma_spin_profiles} for J0846 and Paloma, respectively.

\end{document}